\documentclass{jfm}
\usepackage{graphicx}
\usepackage{epstopdf, epsfig}

\usepackage{ragged2e}                   
\justifying                             
\usepackage{xcolor}						
\usepackage{wasysym}
\usepackage{amssymb}                    
\usepackage[normalem]{ulem}
\usepackage{soul} 

\shorttitle{Air-jet impact craters}
\shortauthor{P. Sonar and H. Katsuragi}

\title{Air-jet impact craters on granular surfaces: a universal scaling}

\author{Prasad Sonar\aff{1}
  \corresp{\email{prasadrsonar@gmail.com}},
 \and Hiroaki Katsuragi\aff{1}}

\affiliation{\aff{1}Department of Earth and Space Science, Osaka University, 1-1 Machikaneyama, Toyonaka, Osaka, 560-0043, JP}

\begin{document}

\maketitle

\begin{abstract}
Craters form as the lander's exhaust interacts with the planetary surfaces. Understanding this phenomenon is imperative to ensure safe landings. We investigate crater morphology, where a turbulent air jet impinges on the granular surfaces. To reveal the fundamental aspect of this phenomenon, systematic experiments are performed with various air-jet velocities, nozzle positions, and grain properties. The resultant crater morphology is characterized by an aspect ratio. We find a universal scaling law in which the aspect ratio is scaled by the dimensionless variable consisting of air velocity at the nozzle, speed of sound in air, nozzle diameter, nozzle-tip distance from the surface, grain diameter, the density of grains, and density of air. The obtained scaling reveals the crossover of the length scales governing crater aspect ratio, providing a useful guideline for ensuring safe landings. Moreover, we report a novel drop-shaped sub-surface cratering phenomenon.
\end{abstract}


\section{Introduction}
People aspire to live on the Moon. To achieve that, we need to develop safe landing mechanisms. Ensuring sustainable and safe landings provides infinite opportunities for further space exploration.
However, attaining a stable landing on a planetary surface stands as a paramount concern in the realm of space immigration. Interaction between the lander’s exhaust and planetary surface is a key component for this problem.
The chosen landing site usually consists of flat and crater-free terrain \citep{toigo2003}.
Nonetheless, the plume ejected from a descending spacecraft has the potential to induce cratering and splashing, leading to a potentially unstable landing scenario.
It is necessary to conduct a thorough assessment of the cratering and regolith grain splashing induced by air jet on the planetary surface.
To explore the inherent nature of this type of phenomenon, it is essential to uncover a universal scaling framework for the granular cratering caused by the impact of air jets.
However, accurately replicating a fully realistic landing scenario is difficult. Consequently, researchers usually perform small-scale laboratory experiments to explore scaling relationships. This study is dedicated to investigating granular cratering resulting from air-jet impact by a small-scale experiment. Through systematic analysis of laboratory-scale experiments, we derive a universal scaling law governing crater morphology. The obtained scaling includes newly defined scaling parameters and offers a novel direction of the research of interaction between fluid and granular matter. In addition, this scaling would allow us to anticipate the potential landing scenarios on various planetary surfaces in future space missions.
%
%
%

Crater formation due to an impact is an extremely complex phenomenon 
\citep{holsapple1982,croft1985,uehara2003,lohse2004,katsuragi2016,yamamoto2017,van2017,prieur2017,allibert2023}.
Jet-impact crater formation further involves the interaction between continuously impinging pressurized fluid and granular surface, which results in bowl-shaped depression. %
The crater shape varies with the properties of the target material and the impinging fluid.
Landing rockets on planetary surfaces and erosion near hydraulic structures are important examples relating to jet-impact-induced cratering \citep{rajaratnam1977,lane2010cratering,badr20141,badr20142,lamarche2015, PNAS2023,metzger2024a}. 
Here, we focus on the cratering phenomena that may cause problems during the landing and re-launching of rockets. Particularly, hardware damage to the rocket body or its sensors can be caused by large amounts of dust kicked up by exhaust plumes. Understanding this type of plume-surface interaction (PSI) process is one of the most important issues in the space engineering field~\citep{donohue2021, Baba:2023, PNAS2023, bajpai2024}. Indeed, dust ejecting up on the sensors was one of the many reasons for recent landing failures on the lunar surface~\citep{witze2023,metzger2024b}. 
\par The granular crater formation caused by the fluid impact has been studied by many researchers~
\citep{rajaratnam1977,lane2010cratering,metzger2009,metzger2011,zhao2013,clark2014,badr20141,badr20142,lamarche2015,badr2016,guleria2020,gong2021,donohue2021,benseghier2023}. 
Various mechanisms reported for crater formation include viscous erosion (VE), diffused gas eruption (DGE), bearing capacity failure (BCF), and diffusion-driven shearing (DDS) \citep{lane2010cratering,metzger2009,metzger2011,kuang2013}. 
The VE is the most common and most investigated \citep{zhao2013,clark2014,badr20141,badr20142,badr2016}. 
The cratering process is controlled by the conditions of the impinging jet and erodible granular bed. During crater erosion, both the crater depth and diameter grow with time and approach the asymptotic values
~\citep{donohue2021,zhao2013,badr20141,PNAS2023}. 
The scaling laws for crater morphology were studied based on the Froude, Shields, and Erosion numbers \citep{gong2021,clark2014,guleria2020}. 
However, these scaling relationships are applicable to only each specific condition. There is a lack of consistency across various experiments, hindering the establishment of a unified understanding among these previous studies. The pursuit of universal scaling relations for granular cratering induced by air-jet impact is a critical issue and intersecting both in fundamental granular physics and space engineering.
In this paper, we propose a unified scaling relation for jet-induced granular cratering, to better understand the PSI process. 
Particularly, new dimensionless parameters are introduced to aim at providing universal explanations for the systematic experimental results.
Besides, we report a novel sub-surface cratering phenomenon as well.%
%
%
\par The rest of the paper is organized as follows: In the next section, the experimental setup, materials, and procedures are introduced. In \S 3, we define the various kinds of craters formed and discuss scaling analysis for crater dimensions using well-known non-dimensional numbers. 
Then, we propose the unified scaling law for the crater's aspect ratio that characterizes the crater morphology. In \S 4, we discuss the advantages of improved scaling relation, further improvements that can be done, and a novel drop-shaped crater that we encounter. Finally, we conclude in \S 5.
%
 
\section{Experiments}
\label{sec:expt}

\begin{figure}
\centering
\includegraphics[scale=0.43]{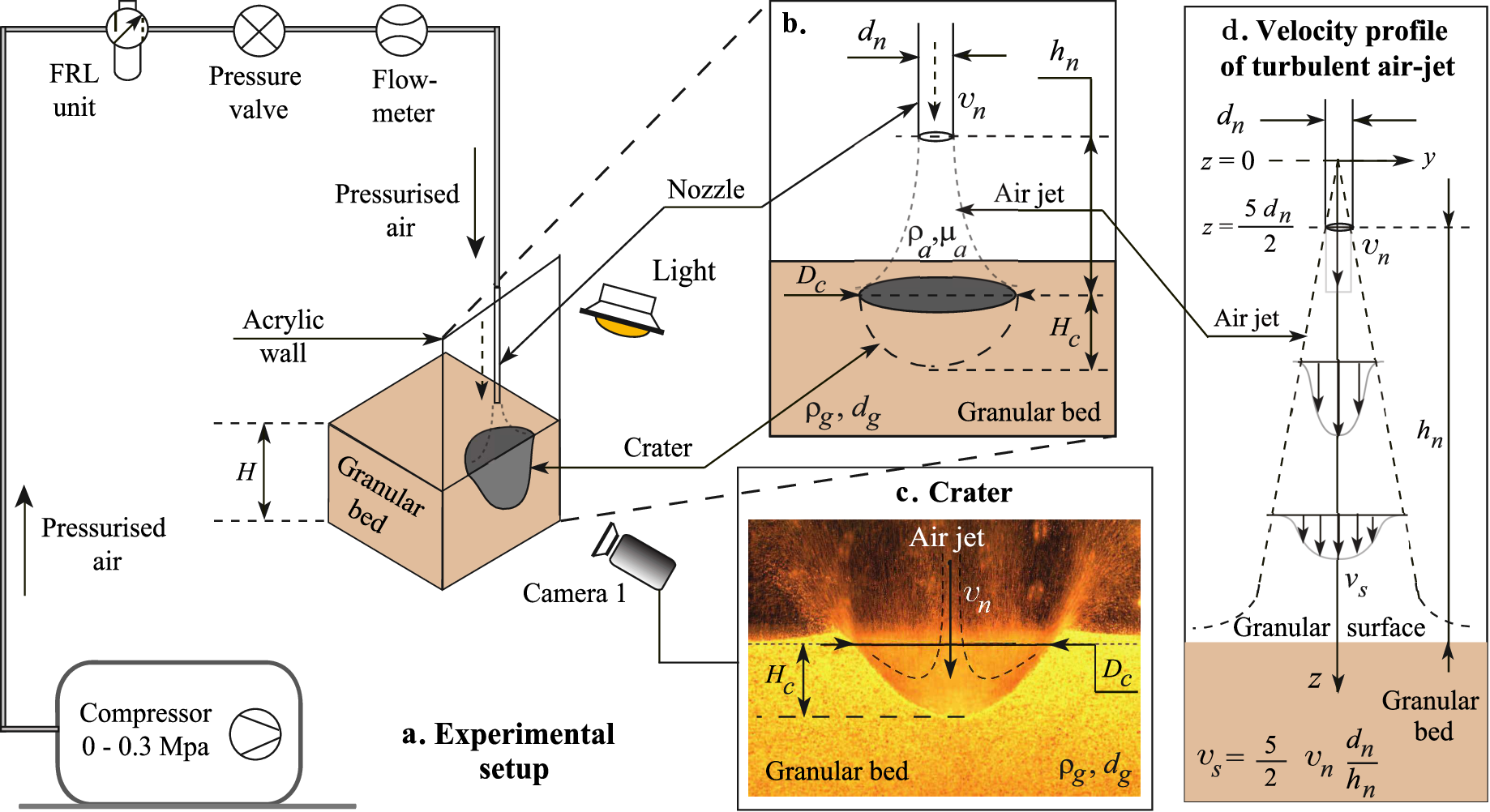}
\caption{(a)~The schematic of experimental setup for air-jet impact experiments in 3D half-space setup, (b)~Magnified view of cratering process, (c)~example image of a crater, and (d)~sketch of the impinging air-jet configuration.}
\label{fig:exptsch}
\end{figure}
\par We systematically perform lab-scale experiments to form air-jet-induced craters. Figure \ref{fig:exptsch} shows the schematic of the experimental setup. We perform experiments in a 3D half-space setup to capture crater images. The rectangular container of inner dimensions, $240 \times 200 \times 70 \, \rm{mm^3}$,
contains the granular material, which serves as a planetary surface simulant.
The compressed air jet, mimicking the nozzle exhaust, is directed vertically downwards along the acrylic wall, keeping a minimum distance of approximately $1 ~\rm{mm}$ between the nozzle and the acrylic wall while minimizing the influence of the wall on the central jet velocity \citep{rajaratnam1977,lamarche2015,schlichting1979,guleria2020}. The formation of boundary layer along the wall is discussed separately in Appendix \ref{appBL}.
%
The pressure-controlled air jet of $0.01$ -- $0.3$~MPa pressure range comes out through a nozzle. We vary nozzle diameter, $d_n$, to control the impinging air-jet velocity, $v_n$. The distance between the granular surface and the nozzle tip, $h_n$, is also varied. The variation in $h_n$ captures the dynamic advancement of the rocket towards the landing surface.
As shown in Fig. \ref{fig:exptsch}(b), the air jet is directed on a granular bed of grain size $d_g$ and true density $\rho_g$.
Upon jet-surface interaction, Fig. \ref{fig:exptsch}(c) captures the formation of the crater of width $D_c$ and depth $H_c$ from the initial surface level (dotted line).
In this study, we focus on the steady crater shapes, i.e., constant $D_c$ and $H_c$ conditions. 
The steady crater shape is immediately developed within a few seconds. Due to a greater tendency to erode, finer and lighter granular material needs more time to reach a steady state than the coarser and heavier grains, respectively.
\begin{table}
  \begin{center}
  \def~{\hphantom{0}}
    \caption{\label{tab:grains} Properties of granular materials.}
    \begin{tabular}{l c c c c c c}
    \textrm{Grains}& $d_g$ $(\rm{mm})$ & $\rho_g$ $(\rm{g/cm^3})$ & Shape & $\varphi_i$ & $\theta \, (^{\circ})$ \\ 
    1. BZ02  & 0.17-0.25 & 2.60 & Sphere & 0.66 & 21.8 $\pm$ 0.7 \\ 
    2. BZ08  & 0.71-0.99 & 2.60 & Sphere & 0.63 & 21.6 $\pm$ 0.7 \\ 
    3. BZ1  & 0.99-1.39  & 2.60 & Sphere & 0.63 & 19.7 $\pm$ 0.3 \\ 
    4. BZ2  & 1.50-2.50  & 2.60 & Sphere & 0.62 & 19.2 $\pm$ 0.2 \\ 
    5. SUS304  & $\simeq 1$  & 7.60 & Cylinder & 0.74  & 30.0 $\pm$ 0.3 \\ 
    6. Sand  & 0.16-0.30  & 2.63 & Irregular & 0.63 & 29.0 $\pm$ 1.6 \\ 
    \end{tabular}
  \end{center}
\end{table}
\begin{table}
  \begin{center}
  \def~{\hphantom{0}}
    \caption{\label{tab:para} Operational range of control parameters.} 
    \begin{tabular}{l c c}
    \textrm{Parameters} & Operational range \\
    1. Air jet velocity, $v_n$  & 19 to 375 $\rm{m/s}$\\
    2. Nozzle diameter, $d_n$  & 2 to 6 $\rm{mm}$\\
    3. Nozzle height, $h_n$  & 10 to 200 $\rm{mm}$\\
    4. Grain's diameter, $d_g$  & 0.16 to 2.5 $\rm{mm}$\\
    5. Grain's density, $\rho_g$  & 2.6 and 7.6 $\rm{g/cm^3}$\\
    6. Grain's shape  & spherical, cylindrical and irregular \\
    \end{tabular}
  \end{center}
\end{table}
\par Table \ref{tab:grains} shows the properties of grains used in the experiments. The initial packing fraction and angle of repose are denoted by $\varphi_i$ and $\theta$, respectively.
The BZ series and SUS304 represent spherical glass beads and cylindrical steel-cut-wire beads, respectively, whereas Sand is irregular-shaped Toyoura sand. The initial volume fraction as mentioned in Table \ref{tab:grains}, would remain approximately the same as uncertainty of the mass in the container is observed to be less than $1\%$ during random sampling measurements.

\par 
The experiments are conducted in the following sequence. We first fill the container by pouring the grains and precisely maintaining the flat surface. The nozzle tip of $d_n$ is placed at $h_n$. %
Then, the cratering induced by the air-jet impact is recorded by the camera (STC-MCCM401U3V) that captures side-view images at $200\, \mathrm{fps}$ with a spatial resolution of 0.058~mm/pixel and $2048 \times 2048$ image size. As shown in Fig. \ref{fig:exptsch}(c), $D_c$ is the width of the crater cavity at the initial horizontal surface level (dotted line) and $H_c$ is the depth of the cavity measured vertically down from the initial surface level. 
During every experiment, an air jet impinges the granular bed for $10 \, \rm{s}$. 
See Tables \ref{tab:PQV3.6} and \ref{tab:PQV6} in Appendix \ref{appexpt} for more details of experimental conditions. 
We consider the average of five measurements taken for each experimental condition. More than 700 total number of experiments are carried out. Table \ref{tab:para} shows the range of control parameters in the experiments. Further details regarding experiments can be found in the Appendix \ref{appexpt}.
%
%

\section{Results}
\label{sec:res}
\begin{figure}
\centering
\includegraphics[scale=0.29]{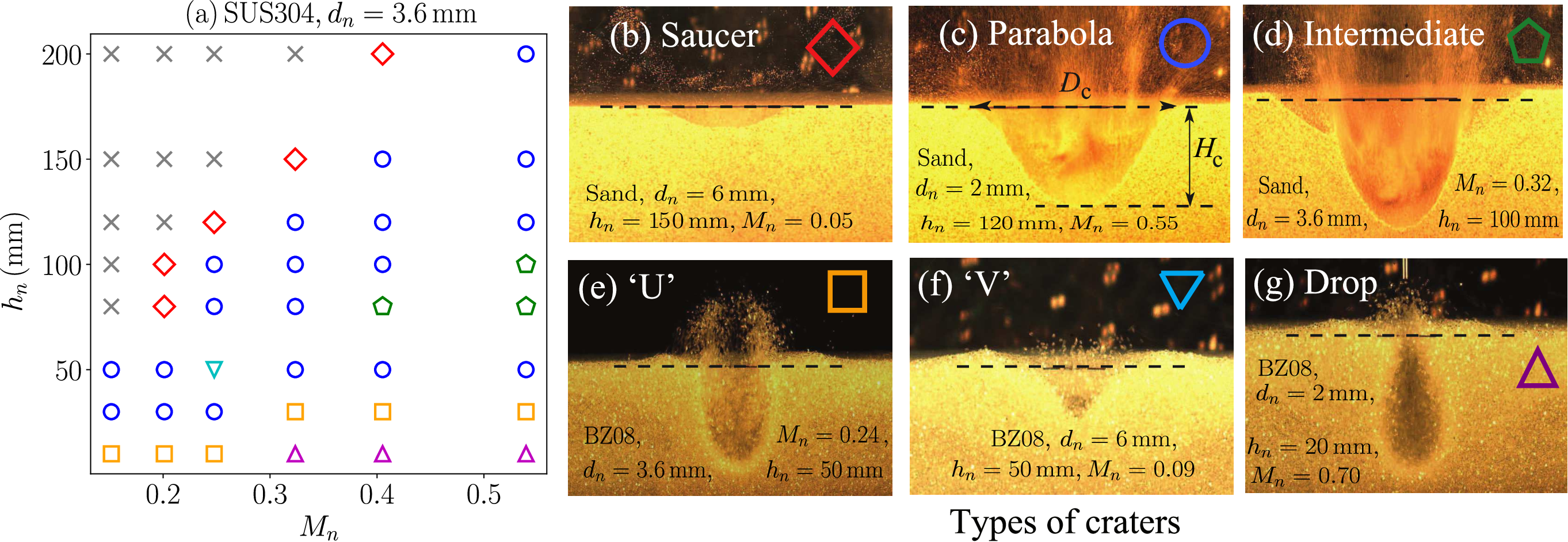}
\caption{(a) Phase diagram shows craters formed when pressurized air jet is directed through the nozzle ($d_n = 3.6 \, \rm{mm}$) on SUS304 grains. Various shapes of craters observed over a range of experimental conditions are: (b) Saucer, (c) parabola, (d) parabola with the intermediate region, (e) U-, (f) V-, and (g) drop-shaped craters. The \textcolor{gray}{$\times$} symbols correspond to the no-crater formations. In the phase diagram, $h_n$ represents the distance from the nozzle tip to the granular surface, and $M_n = v_n/C$ is the dimensionless air-jet velocity at the nozzle, where $C=343$~m/s is the speed of sound in air.}
\label{fig:ph_shapes}
\end{figure}
\par Figure \ref{fig:ph_shapes}(a) shows a phase diagram of various crater types for varying $h_n$ and $M_n$, where $M_n = v_n/C$ is the dimensionless air-jet velocity at the nozzle and $C=343$~m/s is the speed of sound in air. 
As shown in Fig. \ref{fig:ph_shapes}(b--g), we observe six types of craters. Five of these, Fig. \ref{fig:ph_shapes}(b--f), are already documented in the literature \citep{lane2010cratering,clark2014,guleria2020}. %
We reproduce these results over a broader range of parameters, 
i.e., by exploring all the controlling variables mentioned in Table \ref{tab:para} that span a considerable wide range, at least over a decade.
As seen in Fig. \ref{fig:ph_shapes}(a), the \lq parabola\rq \, craters (\textcolor{blue}{$\boldsymbol{\circ}$}) and the parabola with \lq intermediate\rq \, region craters (\textcolor{green}{$\boldsymbol{\pentagon}$}) are observed most frequently within the current parametric range. However, \lq V-shaped\rq \, craters  (\textcolor{cyan}{$\bigtriangledown$}) are formed more frequently for finer grains (see Fig.~\ref{fig:allphases} and \ref{fig:scalephases} in Appendix \ref{appphases} for all phase diagrams and the congregated phase diagram, respectively).
The \lq saucer\rq \, type wide and shallow craters (\textcolor{red}{$\boldsymbol{\diamondsuit}$}) and \lq U-shaped\rq \, narrow and deep craters (\textcolor{orange}{$\boldsymbol{\square}$}) are formed at high and low $h_n$ ranges, respectively.
Moreover, we find a novel crater shape (Fig. \ref{fig:ph_shapes}(g)), \lq drop-shaped\rq \, cavity beneath the granular surface (\textcolor{magenta}{$\boldsymbol{\triangle}$}), which has not been reported in the literature yet. 
While we also observe \lq truncated-shape\rq \, craters due to the size limit of the experimental setup (see Fig.~\ref{fig:images}~(a-b) in Appendix \ref{appimages}), we exclude these truncated craters from the following analysis because the crater's depth $H_c$ cannot be measured for them. The details and movies related to various types of crater formation can be found in the Appendix \ref{appvideos} and the supplementary material (SM).
%
%
\par Now, considering the varied and complex range of craters, we analyze the crater morphology and its governing parameters in detail. 
To obtain the scaling relation, we introduce two dimensionless numbers, $G_n = d_n/h_n$ and $r=\rho_g/\rho_a$, where $\rho_a=1.2$~kg/m$^3$ is the air density. The $G_n$ relates to the velocity profile of the turbulent jet as shown in Fig.~\ref{fig:exptsch}(d). The velocity of the air jet would reduce to $v_s = (5/2)M_n \, G_n \, C$\footnote{For turbulent jets, the universal angle of a diverging cone is approximately $24^{\circ}$. Thus, the initial jet radius and the downstream distance $z$ from the nozzle exit are related by a constant $\tan(12^{\circ})\sim 1/5$. The distance $z$ is counted not from the nozzle exit but from a distance $5d_n/2$ into the nozzle. This point of origin is called the virtual source~\citep{cushman2014}. We use this axisymmetric model for our 3D half-space experimental setup as a first step to investigate the scaling relation.} at the surface of the granular bed~\citep{cushman2014}. By using $r$, a dimensionless number proportional to the dynamic pressure of the air jet, $\rho_a v_n^2$, can be expressed by $(M_n/r^{1/2})^2$. 
From the measured data, we find that $D_c$ and $H_c$ can be scaled as $D_c \sim G_n^{-1}$ and $H_c \sim M_n/r^{1/2}$.
\begin{figure*}
\centering
\includegraphics[scale=0.28]{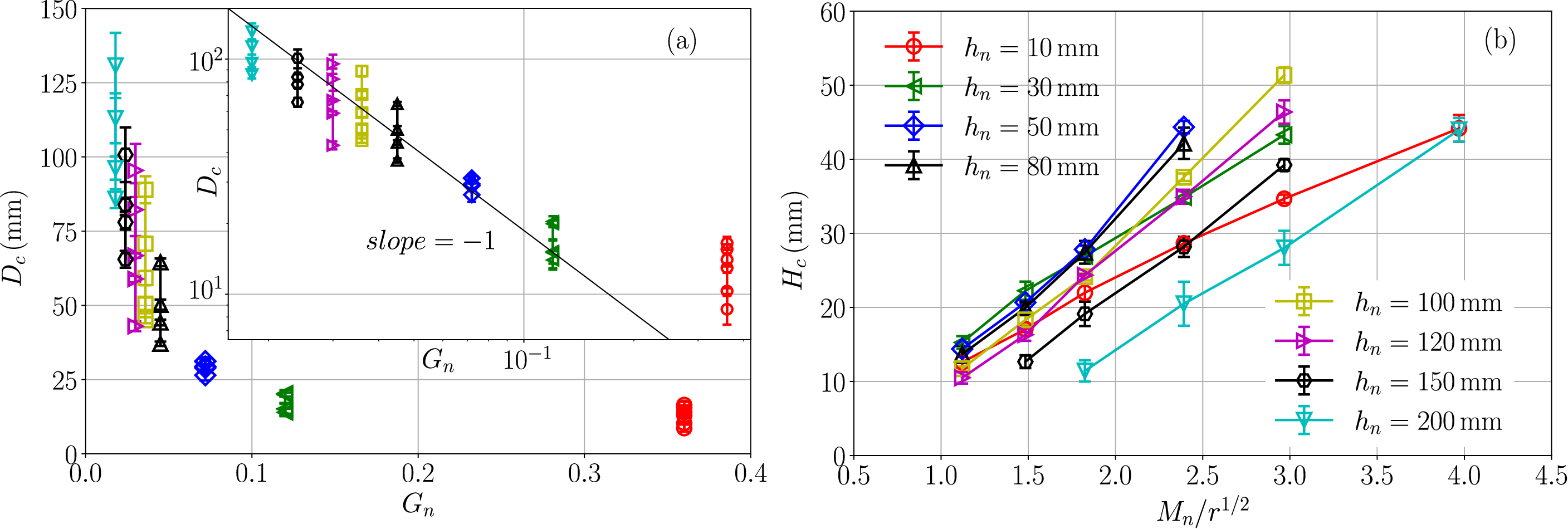}
\caption{The scaling relations among $D_c$, $H_c$, $G_n = d_n/h_n$, $M_n$ and $r=\rho_g/\rho_a$ are presented. Panels (a) and (b) suggest the scaling relations as $D_c \sim G_n^{-1}$ and $H_c \sim (M_n/r^{1/2})^1$, respectively. The inset of (a) shows the plot of the same data on the log-log scale. The data shown in (a) and (b) originate from the experiments performed with BZ1 grains using $d_n = 3.6 \, \mathrm{mm}$.}
\label{fig:DcHcMnhn}
\end{figure*}
Figure~\ref{fig:DcHcMnhn}~(a) and (b) shows the relations $D_c$ vs $G_n$ and $H_c$ vs $M_n/r^{1/2}$, respectively.
Namely, $D_c$ is mainly governed by air-jet geometry and $H_c$ is principally determined by the dynamic pressure of the impinging air jet.
From these relations, we investigate a scaling law for the crater's aspect ratio $R_c=D_c/H_c$.
For the safe rocket landing situation, large-$R_c$ (shallow and flat) cratering causes the ejecta to flow radially outward and is thus preferred over vertically ejected grains that might cause the malfunction of the lander as in the case of deep crater formation. Therefore, $R_c$ is the most important parameter characterizing this type of cratering.
%
%
\begin{figure}
\centering
\includegraphics[scale=0.28]{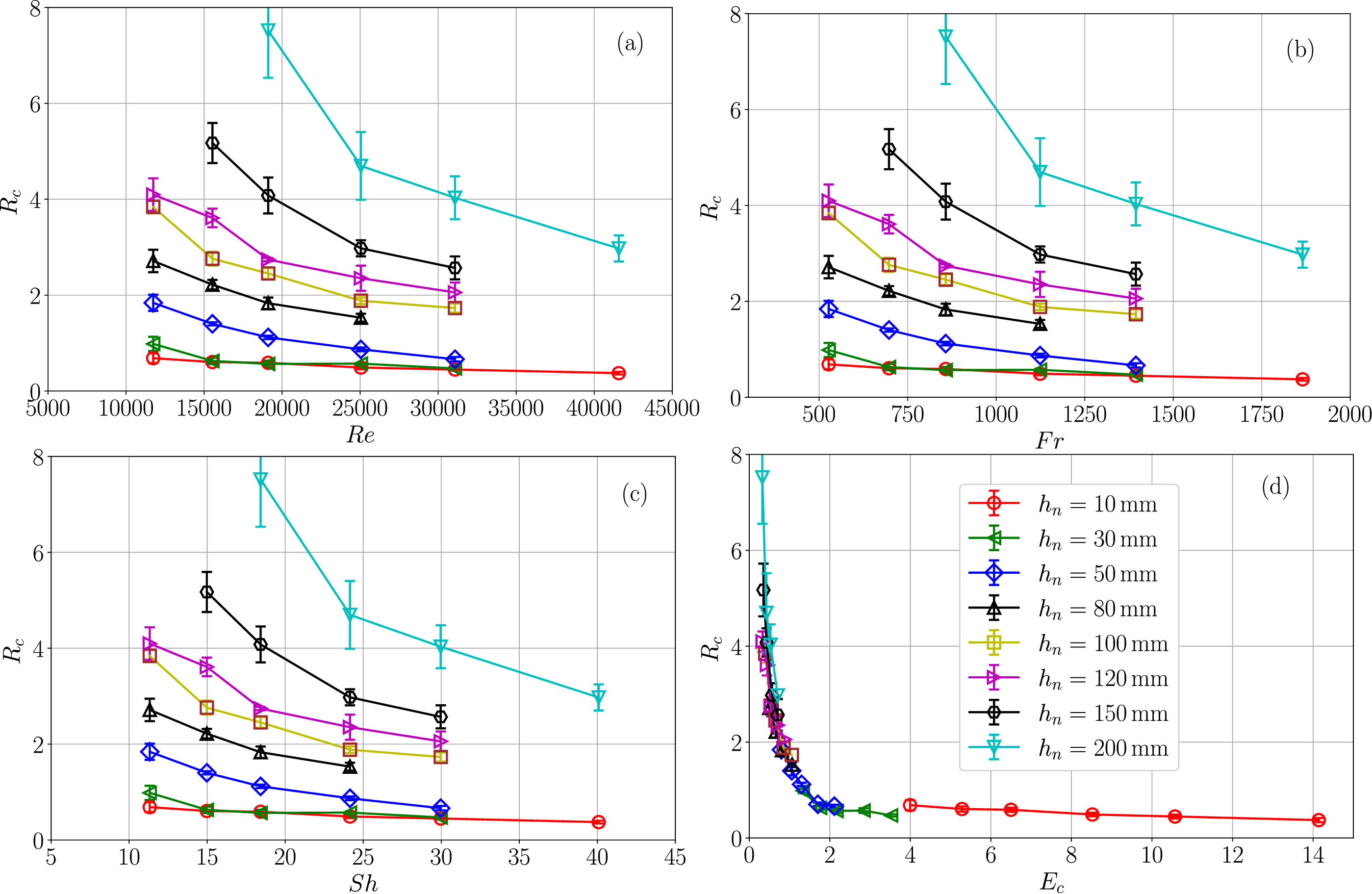}
\caption{Crater's aspect ratio, $R_c$, vs. (a) $Re$, (b) $Fr$, (c) $Sh$, and (d) $E_c$. Experiments are performed for BZ1 and $d_n = 3.6 \, \mathrm{mm}$.}
\label{fig:Rc_numbers}
\end{figure}
\begin{figure} 
\centering
\includegraphics[scale=0.32]{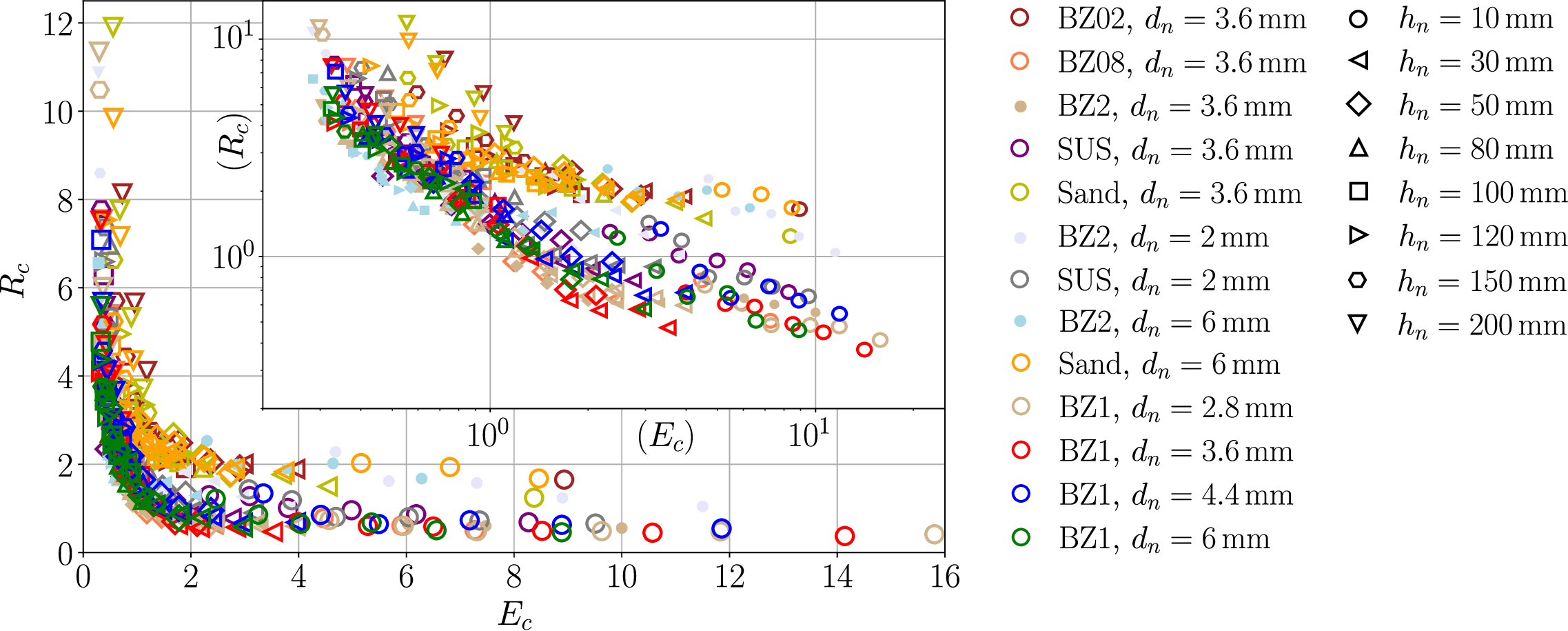}
\caption{The aspect ratio, $R_c$, is plotted as a function of $E_c$. A certain degree of collapse can be observed, but the quality of data collapse is not very good.
Inset shows the plot of the same data on the log-log scale. The shape of symbols represents various $h_n$ values.}
\label{fig:Ec}
\end{figure} 
\par The roles of conventional dimensionless numbers have been thoroughly investigated by researchers \citep{badr2016,guleria2020,gong2021}. They used Froude number $Fr = {v_n}/\sqrt{g d_g}$, Shield's number $Sh = Fr \, \sqrt{{\rho_a}/{(\rho_g - \rho_a)}}$, and Reynolds number $Re = \rho_a v_n d_n/ \mu_a$, where $g$ is the gravitational acceleration and $\mu_a$ is the viscosity of air.
For our case, we observe that the $Re$, $Fr$, and $Sh$ are not sufficient to scale $R_c$ since $h_n$ is not included in these dimensionless numbers; as shown in Fig. \ref{fig:Rc_numbers}(a-c), respectively.
In addition, while studying the temporal evolution of a crater formed by rocket exhaust, Rajaratnam \textit{et al.} introduced an erosion number $E_c =Sh~G_n$, which incorporates various relevant parameters~\citep{rajaratnam1977,donohue2021}. 
However, during an experimental investigation of the growth rate of the crater, Donohue \textit{et al.} indicated that $E_c$ may not accurately characterize the crater formation \citep{donohue2021}. 
Remarkably, focusing on the crater's morphology observed in this paper, the erosion number, $E_c$, indeed shows promising results but only when 
one kind of granular surface is considered (see Fig. \ref{fig:Rc_numbers}(d)). 
Yet, Fig. \ref{fig:Ec} suggests that the data scatters for other experimental parameters.\\
%

%
\begin{figure*}
\centering
\includegraphics[scale=0.32]{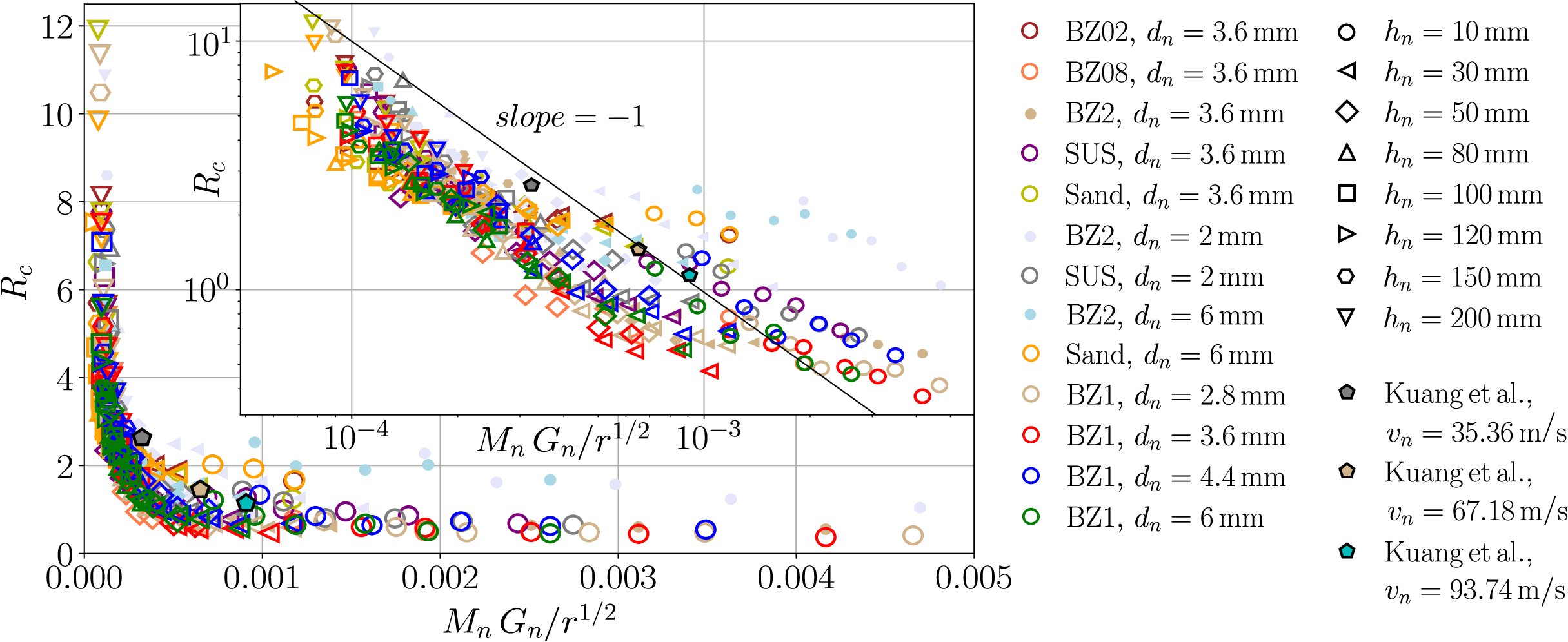}
\caption{The crater aspect ratio $R_c$ of all experiments is plotted as a function of $M_n \, G_n /r^{1/2}$. 
Inset shows the plot of the same data on the log-log scale. The color indicates the target material and $d_n$ conditions. The shape of symbols represents $h_n$ values. The filled circular symbols represent data for $h_n = 10 \, \mathrm{mm}$ of each corresponding color condition.}
\label{fig:MnGnr}
\end{figure*}
%
%
\par Now, we consider a more appropriate choice for a scaling law.
From Fig.~\ref{fig:DcHcMnhn}~(a) and~(b), $D_c\sim G_n^{-1}$ and $H_c\sim M_n/r^{1/2}$ results in the combined scaling $R_c \sim (M_n \, G_n/r^{1/2})^{-1}$.
Figure~\ref{fig:MnGnr} shows the scaling $R_c \sim (M_n \, G_n /r^{1/2})^{-1}$. One can confirm the reasonable data collapse, particularly in small $M_n \, G_n /r^{1/2}$ regime. The obtained scaling is almost consistent with the numerical simulation of air-jet impact onto a granular bed \citep{kuang2013}, filled \textcolor{black}{$\boldsymbol{\pentagon}$} in Fig.~\ref{fig:MnGnr}. 
Although this scaling reasonably collapses the experimental data, we realize the data of small $h_n$ (filled circular symbols) systematically deviate from the scaling.\\
%
%

\begin{figure}
\centering
\includegraphics[scale=0.30]{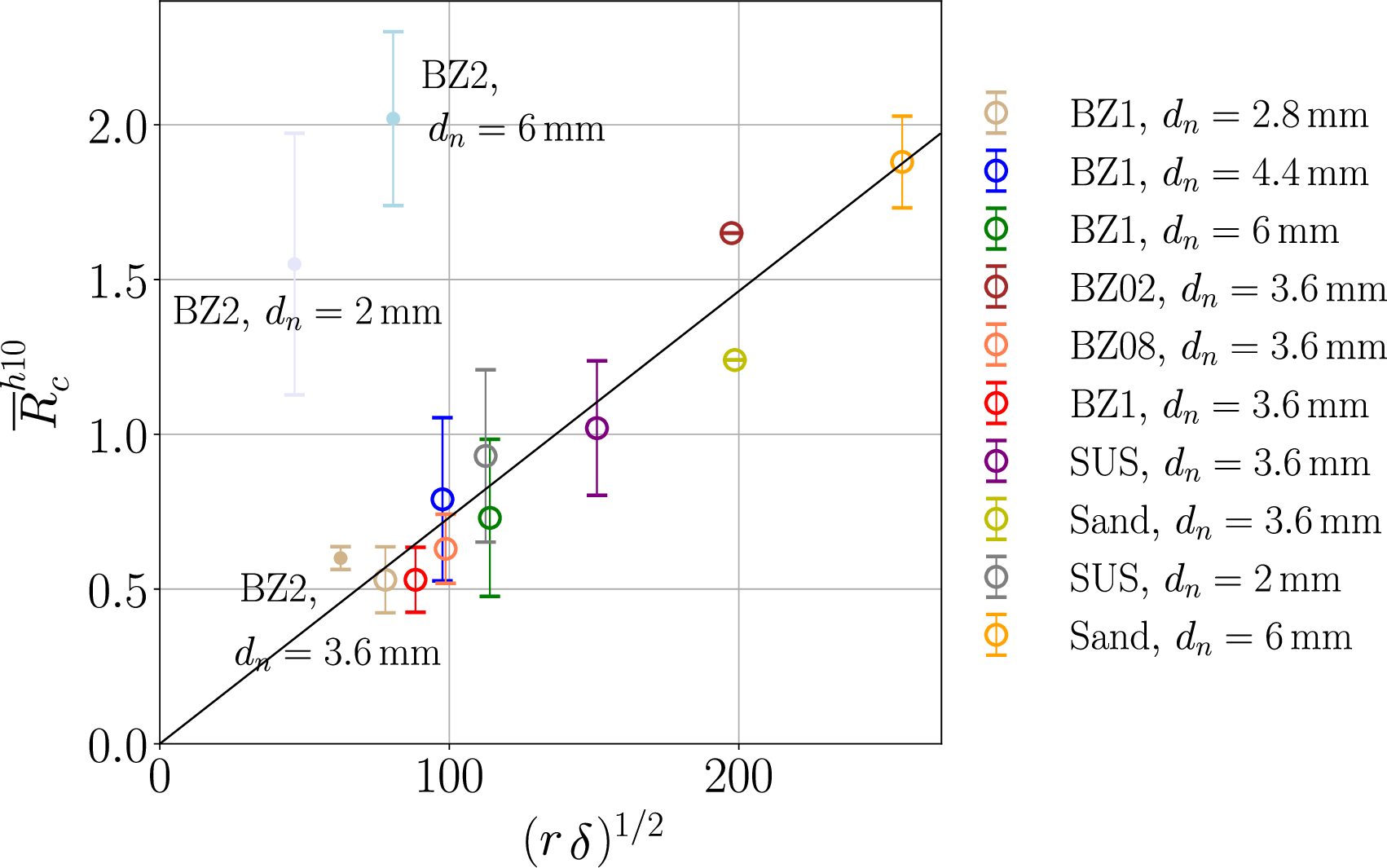}
\caption{The average of aspect ratio $R_c$ at $h_n = 10 \, \mathrm{mm}$, $\overline{R}^{h10}_c$, is plotted as a function of $(r\delta)^{1/2}$.} 
\label{fig:rdelta}
\end{figure}

To obtain better scaling, $R_c$ at the smallest $h_n$($=10$~mm) is further analyzed. We find the average of $R_c$ at $h_n=10$~mm, $\overline{R}^{h10}_c$ is scaled as $\overline{R}^{h10}_c \sim (r \delta)^{1/2}$, where $\delta=d_n/d_g$. 
Figure \ref{fig:rdelta} clearly indicates the scaling $\overline{R}_c^{h10}\sim (r \delta)^{1/2}$ except for the largest grains BZ2 ($d_g=2$~mm).
This scaling in the small $h_n$ regime results from switching of the relevant length scale from $h_n$ to $d_g$, as $h_n$ decreases. Thus, as the lander approaches the surface, the grain size becomes more significant compared to the nozzle height.\\

%

\textcolor{black}{The form of scaling $R_c=(r\delta)^{1/2} = (\rho_g d_n/(\rho_a d_g))^{1/2}$ is slightly puzzling. The density and length scale of air jet and grains are not naturally linked to the crater's aspect ratio. This counter-intuitive correlation could originate from the deformability of the impactor ~\citep{katsuragi2010},} as well as the mixing of the granular substrate and the impactor \citep{nefzaoui2012,zhao2015,zhaomeer2015,zhao2017,van2017}. 
For example, cratering by a droplet impact on a permeable substrate also shows a similar inverse tendency~\citep{katsuragi2010}.
Specifically, a similar density dependence of the crater radius formed by droplet impact was reported. Namely, the larger crater diameter was observed for larger $\rho_g$. 
Recently, Zhao \textit{et al.} performed experiments to study the droplet impact on sand \citep{zhaomeer2015, zhao2017}. It was observed that liquid-grain mixing suppresses droplet spreading and splashing. During the experiments, the packing density of the granular target, grain size, wettability conditions, and the impact velocity are varied. It was found that by increasing the grain size and the wettability conditions, the maximum droplet spreading undergoes a transition from a capillary regime towards a viscous regime. This complex interaction between droplet intruder and granular target creates various crater morphologies \citep{zhao2017}.
These tendencies could be typical difficulties in soft-impact studies.
In our case, high-velocity air can penetrate the shallow layers of the permeable granular substrate with more ease than that of the water droplet. 
The low dynamic viscosity of air, compared to water or other liquids, allows for easier permeation of the granular bed, viewed as a porous medium. This prolonged presence of jet air within the subsurface layers while the crater is being formed may alter the local properties of the granular target during crater formation.
Thus, the empirical scaling obtained here can be used to formalize the scaling better than the previously proposed ones.\\
%
%
\begin{figure}
\centering
\includegraphics[scale=0.5]{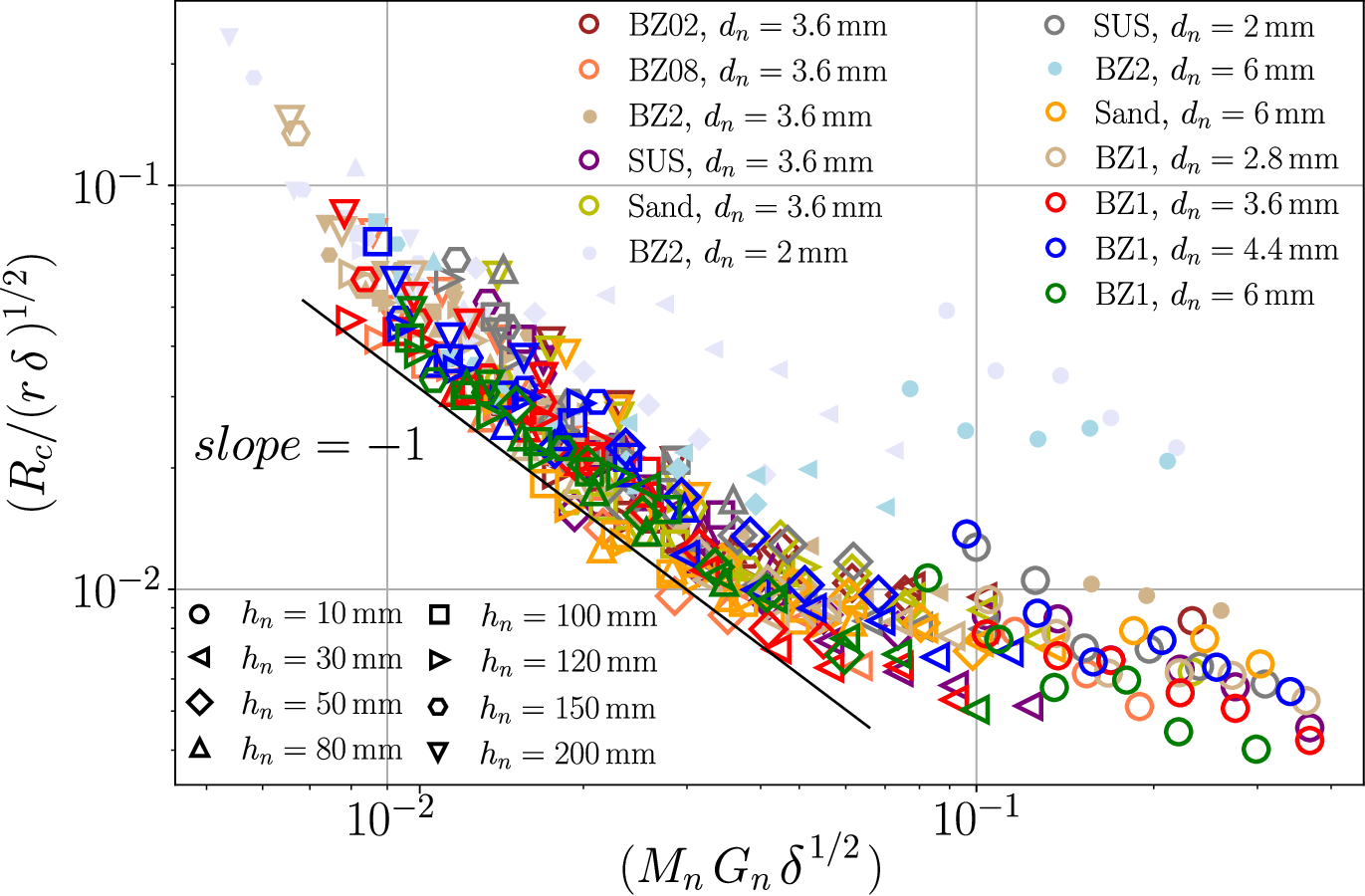}
\caption{The scaling relation of Eq.~(\ref{eq:scaling}), $R_c / (r \delta)^{1/2}=f(M_n G_n \delta^{1/2})$, is presented for all experimental data. Except for the largest grain data (BZ2, $d_g=2$~mm, filled symbols in the plot), all the data obey the scaling. The shape of symbols represents various $h_n$ values.}
\label{fig:RcMnGn}
\end{figure}

Finally, we derive the unified scaling law based on the above-mentioned scaling results. The combined scaling function is written as,
\begin{equation}
\frac{R_c}{(r \delta)^{1/2}} = f\left( M_n G_n \delta^{1/2} \right),
\label{eq:scaling}
\end{equation}  
where the function $f(x)$ satisfies $f(x) \sim x^{-1}$ in the small $x$ regime and $f(x)\simeq$ const.~in the large $x$ regime.
This functional form indeed recovers $R_c \sim (M_nG_n/r^{1/2})^{-1}$ at large $h_n$ and $R_c \sim (r \delta)^{1/2}$ at small $h_n$. In Fig.~\ref{fig:RcMnGn}, the scaling crossover point is $M_n G_n \delta^{1/2} \simeq 10^{-1}$. Therefore, the crossover $h_n$ can be computed as $h_c \simeq 10~M_n \delta^{1/2}d_n$. In such a complex PSI process, capturing a shift in relevant length scale and its crossover point are important takeaways of this paper.

When the grain size is the largest (BZ2 grains), however, the data cannot be scaled (filled symbols in Fig.~\ref{fig:RcMnGn}). This indicates that the cratering dynamics is completely different when $d_g \sim d_n$. Namely, $d_n$ must be sufficiently greater than $d_g$ to safely apply the obtained scaling. This result is informative to provide the lower limit of the nozzle size in the lander design.
%
%
\section{Discussion}
\label{sec:disc}
\par The unified scaling law presented in Fig.~\ref{fig:RcMnGn} (and Eq.~\ref{eq:scaling}) is superior to the previously reported scaling laws that are based on conventional dimensionless numbers. 
\par Moreover, from the fluid mechanics perspective, these scaling relations can be further investigated to study the grain motion and convection currents during the temporal evolution of the different types of craters.
In addition, our result shows the crossover of scaling laws, which may be of interest to researchers studying soft matter physics~\citep{maruoka2023}.
%
%
By using the obtained scaling, $R_c \sim (M_n G_n/r^{1/2})^{-1}$, we can estimate $R_c$ mainly from the jet conditions in the large $h_n$ regime. We do not need detailed information about the landing surface. Having a simple scaling parameter mainly consisting of jet conditions is advantageous to reduce uncertainty in space missions.  Within the approachable distance, we should use the relation $R_c \sim (r \delta)^{1/2}$, i.e., grain size information is necessary in the small $h_n$ regime.

The 3D half-space experimental setup has limitations like the presence of an acrylic wall that forms a boundary layer affecting jet velocity structure. Thus, this setup might break the axisymmetric assumption for the jet. This may affect the surface erosion and the crater dimensions. 
We believe that by addressing such experimental limitations, the current scaling relations (in Fig. \ref{fig:RcMnGn}) may further be improved by reducing data scattering around the scaling.
To consider the realistic rocket landing application, we have to estimate ejection speed, angle, etc. Besides, the effects of gravity and ambient air must be evaluated. In this experiment, dimensionless numbers including the gravity effect ($Fr$, $Sh$, and $E_c$) do not work well to collapse the data (Figs.~\ref{fig:Rc_numbers} and \ref{fig:Ec}). As demonstrated in \citet{Baba:2023}, however, air-jet cratering and the resultant ejection process are significantly affected by the gravity condition. Modification of the obtained scaling and the precise measurement of ejector behavior are the important next steps to consider in the actual landing application. The scaling function developed in this study provides a starting point for further investigations.

As mentioned earlier, we observe a novel drop-shaped subsurface cratering phenomenon on a few occasions, especially when the nozzle tip is very close to the granular surface (see \textcolor{magenta}{$\boldsymbol{\bigtriangleup}$} in Fig.~\ref{fig:ph_shapes}(a)) consisting of coarser/denser grains. In low $h_n$ cases, the turbulent jet impinges the granular surface at high $v_s$.
In the case of coarser grains, the larger pores between the grains allow easier penetration of the high-velocity turbulent air jet. Then, the jet penetrates through the surface and expands beneath. Contrastively, the granular surface made of finer grains is easy to erode. In such a case, the air escapes out while eroding the surface and a drop-shaped cavity cannot be formed. The denser and coarser SUS304 grains are difficult to displace, but allow the trapped air to deform the sub-surface material (see Fig.~\ref{fig:images}(c) in Appendix~\ref{appimages}).

The drop-shaped crater can be attributed to the BCF or DGE mechanisms \citep{metzger2009,alexander1966,scott1968}. However, BCF forms craters with long and narrow cylindrical shapes when pressurized rocket exhaust pushes the soil down into a depression. In contrast, we observed convection currents within the drop-shaped craters with narrow $D_c$ and minor displacement of grains around the drop-cavity as air diffuses into the material.
Thus, we believe, this phenomenon can be related to the diffused gas eruption (DGE)~\citep{scott1968,metzger2009}, which usually causes an annular-ring eruption around the jet. However, we have not seen such eruptions in our small-scale experiments. Further study is necessary to reveal the physical mechanism governing the drop-shaped crater. In addition, this intriguing phenomenon is particularly significant to understanding the rocket launching scenarios from planetary surfaces, where sudden high-speed exhaust thrust needs to be applied on the unknown granular surfaces from a closer range. Moreover, this phenomenon may be of particular importance on small bodies such as the asteroids, which have mostly porous structures.
%
%
\section{Conclusion}

\par We conclude that crater morphology produced by the air-jet impact on the granular surface is primarily governed by the combination of dimensionless air-jet velocity, $M_n = v_n/C$, nozzle geometric factor, $G_n = d_n/h_n$, density ratio between granular target and air jet, $r=\rho_g/\rho_a$, and the diameter ratio between nozzle and grain, $\delta=d_n/d_g$. We obtain the scaling $R_c \sim (M_n G_n/ r^{1/2})^{-1}$, which is useful to estimate the excavation condition by simple scaling parameters, as far as the nozzle tip is away from the granular surface. 
Moreover, we find that within the approachable distance of the granular surface, $R_c$ is governed by $(r \delta)^{1/2}$. This means that the relevant length scale governing crater morphology switches from $h_n$ to $d_g$ by approaching the granular surface.
Finally, the unified scaling function involving both scaling relations is established (Eq.~\ref{eq:scaling}). The obtained scaling is applicable when $d_n > d_g$. 
In addition, we find a novel drop-shaped cratering phenomenon that has potential significance in jet impact physics.
%
%
\appendix

\setcounter{table}{0}
\renewcommand{\thetable}{A\arabic{table}}
\setcounter{figure}{0}
\renewcommand{\thefigure}{A\arabic{figure}}
%

\section{Experimental conditions}\label{app1}
\label{exptcond}
%
%
\subsection{Boundary layer}\label{appBL}
We use a 3D half-space experimental setup that allows us to capture the evolution of the crater and measure the crater's diameter, $D_c$, and depth, $H_c$. However, unlike in the case of the free jet in the complete 3D space, when the turbulent air jet flows parallel to the acrylic wall with one edge of a nozzle close to the wall, it forms a wall jet \citep{rajaratnam1977,launder1983,yue2001,barenblatt2005PNAS,schlichting1979,guleria2020}. A pressure difference causes a jet to deflect towards a boundary that establishes a wall jet at some distance from the nozzle outlet. A wall jet is usually a combination of two distinct layers: an inner boundary layer along the wall and an outer free jet flow layer. These two layers are separated by a mixing layer, where the velocity is close to the maximum \citep{barenblatt2005PNAS}. 
We are interested in a fully developed jet that would hit the granular surface with the maximum velocity.
The wall jet itself is a very complex phenomenon that is being studied to understand its implications \citep{launder1979,launder1983,barenblatt2005PNAS,schlichting1979}. Boundary layer formation along the wall is one such effect. 
For turbulent flows along the wall, the boundary layer thickness \citep{schlichting1979} is given by $\delta_y = 0.37~(h_n/{Re_y}^{1/5})$, where $Re_y = v_n~h_n / \nu$ is Reynolds number and $\nu$ is the kinematic viscosity of air. For the 3D half-space experiments, where the jet is close to the wall, considering the maximum jet length as ours, i.e., $h_n=200~\rm{mm}$, the boundary layer thickness ranges from $1.1$ to $2.3~\rm{mm}$ for the highest and lowest jet velocities, respectively. The maximum value of $\delta_y$ goes down to $1.3~\rm{mm}$ for $h_n=100~\rm{mm}$ at the same lowest velocity.
The reduced velocity at the boundary layer may affect the erosion of the granular surface and, thus, the crater dimensions.
Therefore, a minimum distance is kept between the nozzle and the acrylic wall to ensure that a wall jet is not established immediately at the nozzle exit but at some distance ahead that would help reduce the significant effects on crater formations. This way, we can reduce the unwanted influence of the wall and the inner layer on the core region of the jet as much as possible, especially for large $h_n$ cases.
%
%
%
\subsection{Parametric conditions}\label{appexpt}

As shown in Fig.~1(a) of the main text, the air is compressed and cleaned by the compressor (HG-DC991AL) and regulator (FRL unit), respectively. Then, the clean compressed air enters the solenoid valve (CKD, EXA-C6-02C-3), which controls the air pressure. Next, the flow meter (HoribaStec, MF-FP 10NH06-500-AI-ANV3M) measures the flow rate, $Q$, up to $50$~LPM. Using a linear relationship between $Q$ and pressure, $P$, we extrapolate for $Q$ beyond its original observation range at higher $P$.
\begin{table}
  \begin{center}
    \caption{\label{tab:PQV3.6} {Air-jet parameters at $d_n = 3.6 \, \mathrm{mm}$.}}
    \begin{tabular}{l c c c c c c}
    \textrm{No.} & $P_s$ $(\rm{MPa})$ & $P$ $(\rm{MPa})$ & $Q$ $(\rm{LPM})$ & $v_n$ $\rm(m/s)$ & $M_n = v_n/C$ & $Re = \rho_a v_n d_n/ \mu_a$\\
	1  & 0.30  & 0.25  & 110.5 $\pm$ 1.2  & 184.7 $\pm$ 2.0  & 0.53 & 41573 \\
	2  & 0.20  & 0.18  & 82.6   $\pm$ 0.6  & 138.1 $\pm$ 1.1  & 0.40 & 31077 \\
	3  & 0.15  & 0.13  & 66.6   $\pm$ 0.4  & 111.3 $\pm$ 0.7  & 0.32 & 25048 \\
	4  & 0.10  & 0.09  & 50.8   $\pm$ 0.4  & 84.9  $\pm$ 0.6  & 0.24 & 19108 \\
	5  & 0.07  & 0.06  & 41.3   $\pm$ 0.2  & 69.0  $\pm$ 0.3  & 0.20 & 15536 \\
	6  & 0.04  & 0.03  & 31.2   $\pm$ 0.2  & 52.1  $\pm$ 0.3  & 0.15 & 11741 \\
    \end{tabular}
    \end{center}
\end{table}
\begin{table}
  \begin{center}
    \caption{\label{tab:PQV6} {Air-jet parameters at $d_n = 6 \, \mathrm{mm}$.}}
    \begin{tabular}{l c c c c c c}
    \textrm{No.} & $P_s$ $(\rm{MPa})$ & $P$ $(\rm{MPa})$ & $Q$ $(\rm{LPM})$ & $v_n$ $\rm(m/s)$ & $M_n   = v_n/C$ & $Re = \rho_a v_n d_n/ \mu_a$\\
	1  & 0.30  & 0.26  & 115.8 $\pm$ 2.9  & 69.6 $\pm$ 1.7  & 0.20 & 26123 \\
	2  & 0.20  & 0.18  & 85.4   $\pm$ 1.1  & 51.4 $\pm$ 0.7  & 0.14 &  19284 \\
	3  & 0.15  & 0.13  & 69.7   $\pm$ 0.8  & 41.9 $\pm$ 0.5  & 0.12 & 15742 \\
	4  & 0.10  & 0.09  & 52.4   $\pm$ 0.4  & 31.6  $\pm$ 0.2  & 0.09 & 11865 \\
	5  & 0.07  & 0.06  & 42.3   $\pm$ 0.3  & 25.4  $\pm$ 0.2  & 0.07 & 9951 \\
	6  & 0.04  & 0.03  & 32.0   $\pm$ 0.07 & 19.2 $\pm$ 0.04  & 0.05 & 7236 \\
    \end{tabular}
    \end{center}
\end{table}
\par
\begin{figure}
\centering
\includegraphics[scale=0.24]{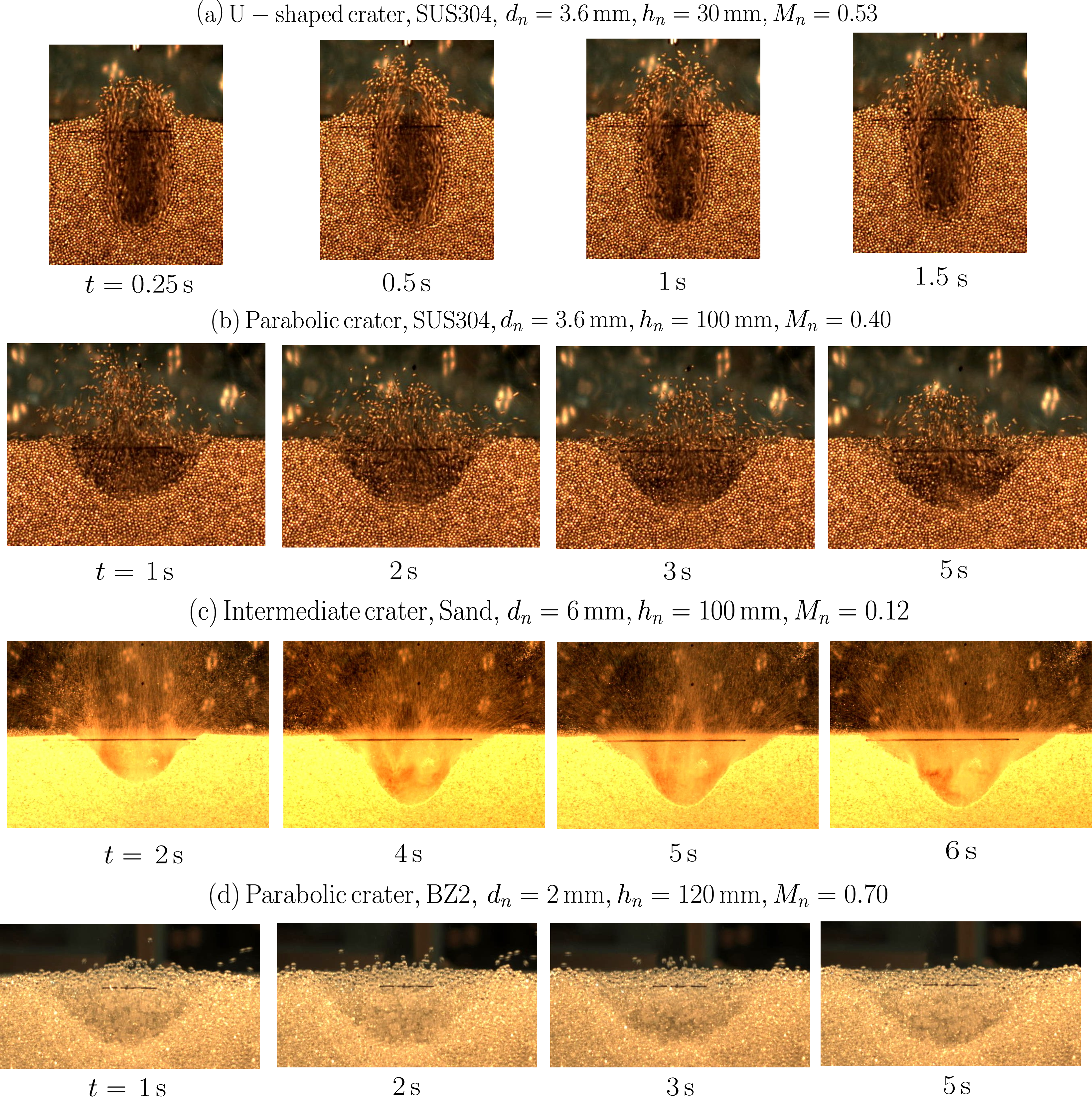}
\caption{The timeline for various craters to achieve a steady state.}
\label{fig:steady}
\end{figure}

 Tables \ref{tab:PQV3.6} and \ref{tab:PQV6} show six different conditions used to perform the experiments when $d_n = 3.6 \, \rm{mm}$ and $6 \, \rm{mm}$, respectively. Due to the limited compressor capacity ($36 \, \rm{L}$), it is difficult to maintain constant pressure while a crater is being formed. All experiments are started at pressure $P_s$ and last for $10 \, \rm{s}$.
As shown in Fig. \ref{fig:steady}, the steady crater shape is immediately developed within a few seconds. As shown various craters reach steady states at different times based on the parametric conditions.
Considering the nozzle diameter from $6 \, \rm{mm}$ to $2 \, \rm{mm}$, the total time required to drop the pressure from $0.3 \, \rm{MPa}$ to $0.01 \, \rm{MPa}$ varies from $90 \, \rm{s}$ to $120 \, \rm{s}$, respectively. The pressure range is determined by the controllable range of the solenoid valve and the flow meter. Thus, within this available time range, we perform experiments over six different air-pressure conditions. During this time, the average quantities of pressure $P$, mass flow rate $Q$, and $v_n$ are evaluated, with corresponding dimensionless air-jet velocity at the nozzle (Mach number), $M_n = v_n/C$ and Reynolds number, $Re = \rho_a v_n d_n/ \mu_a$, where $C$ is the speed of sound in air, $d_n$ is nozzle diameter, $\rho_a$ and $\mu_a$ are density and viscosity of air, respectively. We use the speed of sound in air, $C = 343 \, \mathrm{m/s}$, to non-dimensionalise the air-jet velocity, $v_n$.
%
%
%
\section{Phase diagrams}\label{appphases}

\begin{figure}
\centering
\includegraphics[scale=0.25]{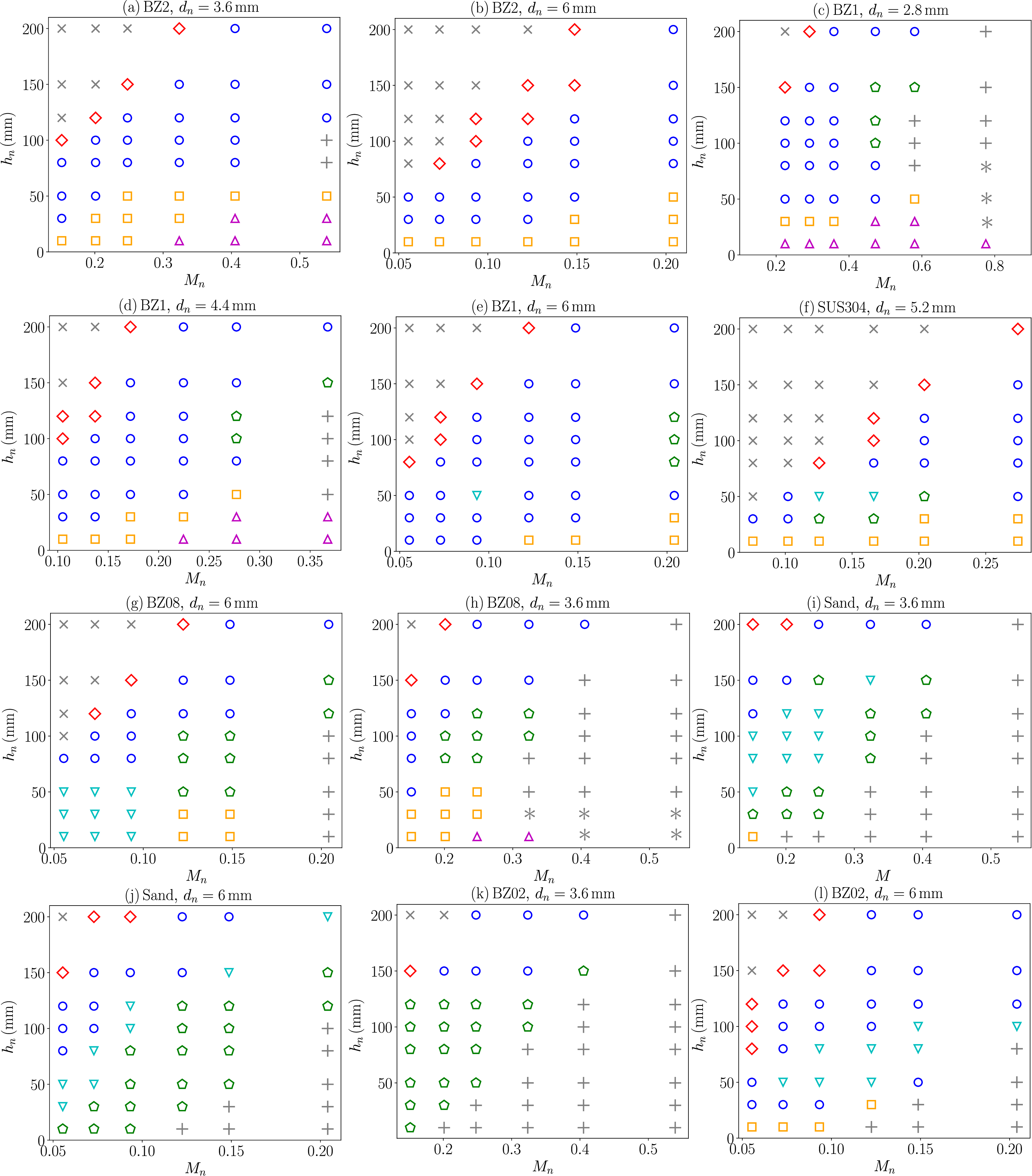}
\caption{Phase diagrams for the experiments performed over a range of material and controlling parameters. Various shapes of craters observed are: Saucer (\textcolor{red}{$\boldsymbol{\diamondsuit}$}), parabola (\textcolor{blue}{$\boldsymbol{\circ}$}), parabola with intermediate region  (\textcolor{green}{$\boldsymbol{\pentagon}$}), U-shaped (\textcolor{orange}{$\boldsymbol{\square}$}), V-shaped (\textcolor{cyan}{$\bigtriangledown$}), and drop-shaped crater (\textcolor{magenta}{$\boldsymbol{\bigtriangleup}$}). The \textcolor{gray}{$+$}, \textcolor{gray}{$*$} and  \textcolor{gray}{$\times$} symbols correspond to the truncated parabolic craters, truncated drop-shaped craters and no-crater formations, respectively. Example images of parabolic and drop-shaped truncated craters can be found in Fig.~\ref{fig:images}~(a,b).
}
\label{fig:allphases}
\end{figure}

%
Figure \ref{fig:allphases} (a-l) shows the phase diagrams of experiments performed with various combinations of granular materials and nozzle parameters $d_n$ and $h_n$. In general, the \lq saucer-shaped\rq \, wide and shallow craters (\textcolor{red}{$\boldsymbol{\diamondsuit}$}) and \lq U\rq \,-shaped narrow and deep craters (\textcolor{orange}{$\boldsymbol{\square}$}) are formed at higher and lower values of $h_n$, respectively. The \lq parabola-shaped\rq \, crater (\textcolor{blue}{$\boldsymbol{\circ}$}) and the parabola with an \lq intermediate\rq \, 
region crater (\textcolor{green}{$\boldsymbol{\pentagon}$}) are observed most frequently within the parametric range. However, \lq V-shaped\rq \, craters (\textcolor{cyan}{$\bigtriangledown$}) are formed more frequently for finer grains. For finer grains, truncated craters ($+$) and truncated drop-shaped craters (\textcolor{gray}{$*$}) are observed at high velocities, which corresponds to the large $M_n$ regime. For large $h_n$ and small $M_n$, the air jet is not able to erode the granular surface to form a crater (represented by $\times$).
\par
\begin{figure}
\centering
\includegraphics[scale=0.42]{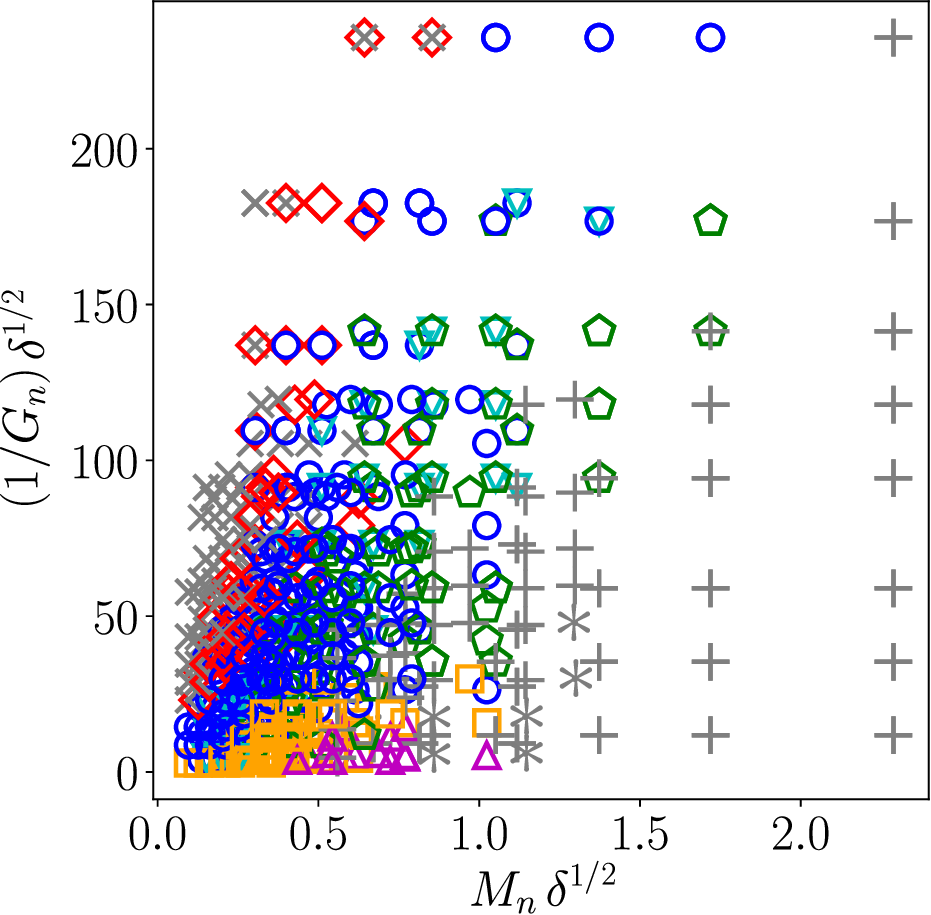}
\caption{Congregated phase diagram. 
Various shapes of craters observed are: Saucer (\textcolor{red}{$\boldsymbol{\diamondsuit}$}), parabola (\textcolor{blue}{$\boldsymbol{\circ}$}), parabola with intermediate region  (\textcolor{green}{$\boldsymbol{\pentagon}$}), U-shaped (\textcolor{orange}{$\boldsymbol{\square}$}), V-shaped (filled \textcolor{cyan}{$\bigtriangledown$}), and drop-shaped crater (\textcolor{magenta}{$\boldsymbol{\bigtriangleup}$}). The \textcolor{gray}{$+$}, \textcolor{gray}{$*$} and  \textcolor{gray}{$\times$} symbols correspond to the truncated parabolic craters, truncated drop-shaped craters and no-crater formations, respectively. In the congregated phase diagram, $\delta=d_n/d_g$, where $d_n$ and $d_g$ are nozzle and grain diameters, respectively.}
\label{fig:scalephases}
\end{figure}

 Figure \ref{fig:scalephases} shows congregated data incorporating all phase diagrams presented earlier in the parametric space of $(1/G_n)  \, \delta^{1/2}$ and $M_n \, \delta^{1/2}$. 
While one can confirm the clustering tendencies of the same symbols, it is difficult to define clear borders among crater shapes.
However, this information may assist in preplanning the lander's descent based on the possibility of forming a minimal-risk crater shape. This involves controlling exhaust velocity at the nozzle at different heights above the surface. This would also allow the lander to estimate the time and amount of dust splashing, which may lead to potential malfunction.\\
%

%
\section{Crater images}\label{appimages}

\begin{figure}
\centering
\includegraphics[scale=0.45]{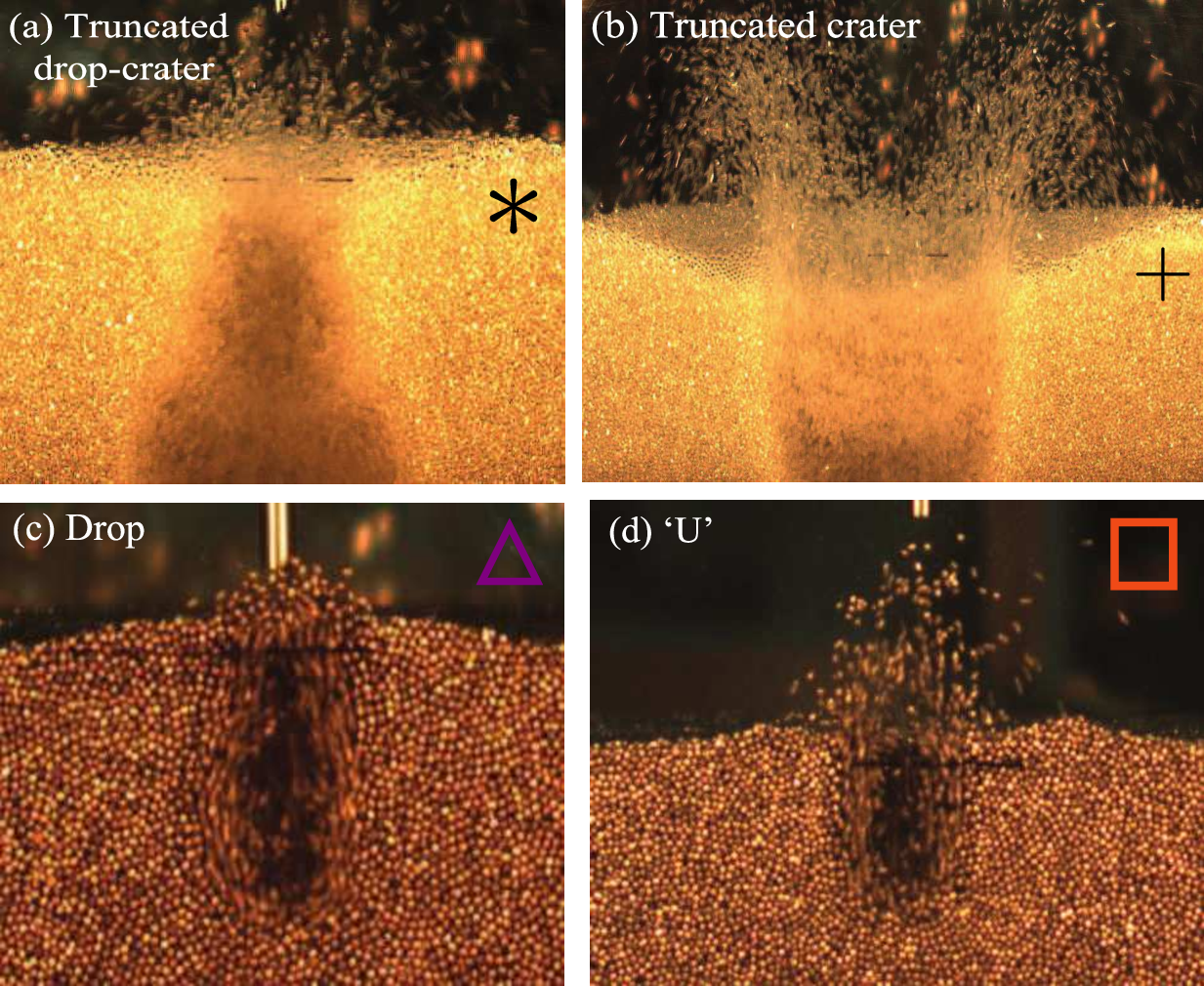}
\caption{
(a) truncated drop-shaped-crater (BZ1, $d_n = 2.8 \, \mathrm{mm}, \, h_n = 30 \, \mathrm{mm}, \, \mathrm{and}, \, M_n = 0.77$), (b) truncated crater (BZ1, $d_n = 2.8 \, \mathrm{mm}, \, h_n = 100 \, \mathrm{mm}, \, \mathrm{and}, \, M_n = 0.77$), (c) drop-shaped crater (SUS304, $d_n = 3.6 \, \mathrm{mm}, \, h_n = 10 \, \mathrm{mm}, \, \mathrm{and}, \, M_n = 0.53$), and (d) \lq U\rq \, shaped crater (SUS304, $d_n = 2 \, \mathrm{mm}, \, h_n = 30 \, \mathrm{mm}, \, \mathrm{and}, \, M_n = 0.7$).}
\label{fig:images}
\end{figure}

Figure \ref{fig:images} shows additional experimental images of crater shapes. Figure \ref{fig:images}~(a) shows a truncated drop-shaped crater ($*$),
 where the air jet is strong enough to go through the granular bed and interact with the base of the container. Here, the nozzle position is close to the surface. Thus, the high-velocity turbulent air jet passes through and expands beneath the surface. We call it a truncated drop-shaped crater, as the crater would have assumed a drop-shape for a thicker bed thickness.
 However, if the nozzle is positioned farther from the surface, it causes high erosion at the surface. Thus, the air jet forms the shape of a truncated crater ($+$) as shown in Figure \ref{fig:images}~(b). We do not investigate the truncated crater phenomenon here because the system size clearly affects the result and it is impossible to measure $H_c$. 
 The data used in this article is only for the complete i.e., non-truncated craters, where the granular material below the crater shows negligible displacement, and thus, such craters are not affected by the container's base. We experimentally confirm that the crater shape is almost independent of the thickness of the target granular layer in $50$ -- $80$~mm depth range.
 Figures \ref{fig:images}~(c) and (d) show \lq drop-shaped\rq \, and \lq U\rq \, shaped craters, respectively, formed with SUS304 material, which has a mass density almost three times greater than that of BZ1 grains.\\
%
\section{Crater videos}\label{appvideos}
%
As listed in Table~A3, we append short movies corresponding to the images of crater formations included in the manuscript (Fig. 1(c) and Fig. 2(b-g)) and Appendix ~\ref{appimages} (Fig. \ref{fig:images}(a-d)) in SM. Table~A3 includes 
Movie names individually describing the type of crater formed under specified experimental conditions.
\begin{table}\label{tab:videos}
  \begin{center}
    \caption{{List of movies corresponding to the crater images discussed in the main text \& in Appendices.}}
    \begin{tabular}{@{}lllllll@{}}
    \textrm {Name} & {Reference} & {Crater type} & {Material} & $d_n$ $(\rm{mm})$ & $h_n$ $(\rm{mm})$ & $M_n$\\
	Movie 1  & Fig. 1(c) & Parabola (\textcolor{blue}{$\boldsymbol{\circ}$}) & Sand & 3.6 & 150 & 0.40 \\
	Movie 2  & Fig. 2(b) & Saucer (\textcolor{red}{$\boldsymbol{\diamondsuit}$}) & Sand & 6.0 & 150 & 0.05 \\
	Movie 3 & Fig. 2(c) & Parabola (\textcolor{blue}{$\boldsymbol{\circ}$}) & Sand & 6.0 & 120 & 0.07 \\
	Movie 4  & Fig. 2(d) & Intermediate (\textcolor{green}{$\boldsymbol{\pentagon}$}) & Sand & 3.6 & 100 & 0.32 \\
	Movie 5 & Fig. 2(e) & U (\textcolor{orange}{$\boldsymbol{\square}$}) & BZ08 & 3.6 & 50 & 0.25 \\
	Movie 6  & Fig. 2(f) & V (\textcolor{cyan}{$\bigtriangledown$}) & BZ08 & 6.0 & 50 & 0.07 \\
	Movie 7  & Fig. 2(g) & Drop (\textcolor{magenta}{$\boldsymbol{\triangle}$}) & BZ08 & 2.0 & 20 & 0.55 \\
	Movie 8 & Fig. A4(a) & Truncated drop (\textcolor{gray}{$*$}) & BZ1 & 2.8 & 30 & 0.77 \\
	Movie 9 & Fig. A4(b) & Truncated ($+$) & BZ1 & 2.8 & 100 & 0.77 \\
	Movie 10 & Fig. A4(c) & Drop (\textcolor{magenta}{$\boldsymbol{\triangle}$}) & SUS & 3.6 & 10 & 0.53 \\
	Movie 11 & Fig. A4(d) & U (\textcolor{orange}{$\boldsymbol{\square}$}) & SUS & 2.0 & 30 & 0.70 \\
    \end{tabular}
    \end{center}
\end{table}\\
%
%

This work was supported by the JSPS KAKENHI, Grant No. JP18H03679 and JP24H00196, and JSPS-DST Bilateral Program, Grant No. JPSJBP120227710.

%
%
\bibliographystyle{jfm}
\bibliography{JFM_ref_2024}

\begin{thebibliography}{45}
\expandafter\ifx\csname natexlab\endcsname\relax\def\natexlab#1{#1}\fi
\def\au#1{#1} \def\ed#1{#1} \def\yr#1{#1}\def\at#1{#1}\def\jt#1{\textit{#1}}
  \def\bt#1{#1}\def\bvol#1{\textbf{#1}} \def\vol#1{#1} \def\pg#1{#1}
  \def\publ#1{#1}\def\arxiv#1{#1}\def\org#1{#1}\def\st#1{\textit{#1}}

\bibitem[Alexander {\em et~al.\/}(1966)Alexander, Roberds \&
  Scott]{alexander1966}
{\sc \au{Alexander, J.~D.}, \au{Roberds, W.~M.} \& \au{Scott, R.~F.}} \yr{1966}
   \bt{Soil erosion by landing rockets final report}.  \org{{\em Tech.
  Rep.\/}}.

\bibitem[Allibert {\em et~al.\/}(2023)Allibert, Landeau, R{\"o}hlen, Maller,
  Nakajima \& W{\"u}nnemann]{allibert2023}
{\sc \au{Allibert, L.}, \au{Landeau, M.}, \au{R{\"o}hlen, R.}, \au{Maller, A.},
  \au{Nakajima, M.} \& \au{W{\"u}nnemann, K.}} \yr{2023}  \at{Planetary
  impacts: Scaling of crater depth from subsonic to supersonic conditions}.
  \jt{Journal of Geophysical Research: Planets}  \bvol{128}~(8),
  \pg{e2023JE007823}.

\bibitem[Baba {\em et~al.\/}(2023)Baba, Okita, Watanabe, Maru, Sawai, Mori \&
  Fujita]{Baba:2023}
{\sc \au{Baba, M.}, \au{Okita, S.}, \au{Watanabe, K.}, \au{Maru, Y.},
  \au{Sawai, S.}, \au{Mori, O.} \& \au{Fujita, K.}} \yr{2023}
  \at{{Microgravity Experiment using Drop Tower and CFD-DEM Coupled Simulation
  about Plume-Surface Interaction}}.  \jt{AIAA SCITECH 2023 Forum} .

\bibitem[Badr {\em et~al.\/}(2014{\natexlab{{\em a\/}}})Badr, Gauthier \&
  Gondret]{badr20141}
{\sc \au{Badr, S.}, \au{Gauthier, G.} \& \au{Gondret, P.}}
  \yr{2014{\natexlab{{\em a\/}}}} Erosion of granular bed by a normal jet.
  \bt{In {\em APS Division of Fluid Dynamics Meeting Abstracts\/}},  \pg{pp.
  R17--002}.

\bibitem[Badr {\em et~al.\/}(2014{\natexlab{{\em b\/}}})Badr, Gauthier \&
  Gondret]{badr20142}
{\sc \au{Badr, S.}, \au{Gauthier, G.} \& \au{Gondret, P.}}
  \yr{2014{\natexlab{{\em b\/}}}}  \at{Erosion threshold of a liquid immersed
  granular bed by an impinging plane liquid jet}.  \jt{Phys. Fluids}
  \bvol{26}~(2),  \pg{023302}.

\bibitem[Badr {\em et~al.\/}(2016)Badr, Gauthier \& Gondret]{badr2016}
{\sc \au{Badr, S.}, \au{Gauthier, G.} \& \au{Gondret, P.}} \yr{2016}
  \at{Crater jet morphology}.  \jt{Phys. Fluids}  \bvol{28}~(3),  \pg{033305}.

\bibitem[Bajpai {\em et~al.\/}(2024)Bajpai, Bhateja \& Kumar]{bajpai2024}
{\sc \au{Bajpai, A}, \au{Bhateja, A} \& \au{Kumar, R}} \yr{2024}
  \at{Plume-surface interaction during lunar landing using a two-way coupled
  dsmc-dem approach}.  \jt{Phys. Rev. Fluids}  \bvol{9}~(2),  \pg{024306}.

\bibitem[Barenblatt {\em et~al.\/}(2005)Barenblatt, Chorin \&
  Prostokishin]{barenblatt2005PNAS}
{\sc \au{Barenblatt, G.~I.}, \au{Chorin, A.~J.} \& \au{Prostokishin, V.~M.}}
  \yr{2005}  \at{The turbulent wall jet: A triple-layered structure and
  incomplete similarity}.  \jt{Proceedings of the National Academy of Sciences}
   \bvol{102}~(25),  \pg{8850--8853}.

\bibitem[Benseghier {\em et~al.\/}(2023)Benseghier, Luu, Cu{\'e}llar, Bonelli
  \& Philippe]{benseghier2023}
{\sc \au{Benseghier, Z.}, \au{Luu, L.~H.}, \au{Cu{\'e}llar, P.}, \au{Bonelli,
  S.} \& \au{Philippe, P.}} \yr{2023}  \at{On the erosion of cohesive granular
  soils by a submerged jet: a numerical approach}.  \jt{Granular Matter}
  \bvol{25}~(1),  \pg{8}.

\bibitem[Clark \& Behringer(2014)]{clark2014}
{\sc \au{Clark, A.~H.} \& \au{Behringer, R.~P.}} \yr{2014}  \at{Jet-induced 2-d
  crater formation with horizontal symmetry breaking}.  \jt{Granular Matter}
  \bvol{16}~(4),  \pg{433--440}.

\bibitem[Croft(1985)]{croft1985}
{\sc \au{Croft, S.~K.}} \yr{1985}  \at{The scaling of complex craters}.
  \jt{Journal of Geophysical Research: Solid Earth}  \bvol{90}~(S02),
  \pg{C828--C842}.

\bibitem[Cushman-Roisin(2014)]{cushman2014}
{\sc \au{Cushman-Roisin, B.}} \yr{2014}  \at{Chapter 9: Turbulent jets}.
  \jt{Environmental Fluid Mechanics, Dartmouth College, Thayer School of
  Engineering}  \pg{pp. 153--161}.

\bibitem[Donohue {\em et~al.\/}(2021)Donohue, Metzger \& Immer]{donohue2021}
{\sc \au{Donohue, C.~M}, \au{Metzger, P.~T} \& \au{Immer, C.~D.}} \yr{2021}
  \at{Empirical scaling laws of rocket exhaust cratering}.  \jt{arXiv preprint
  arXiv:2104.05176} .

\bibitem[Gong {\em et~al.\/}(2021)Gong, Azadi, Gans, Gondret \&
  Sauret]{gong2021}
{\sc \au{Gong, M.}, \au{Azadi, S.}, \au{Gans, A.}, \au{Gondret, P.} \&
  \au{Sauret, A.}} \yr{2021} Erosion of a cohesive granular material by an
  impinging turbulent jet.  \bt{In {\em EPJ Web of Conferences\/}}, ,
  \vol{vol. 249},  \pg{p. 08011}. EDP Sciences.

\bibitem[Gorman {\em et~al.\/}(2023)Gorman, Rubio, Diaz-Lopez, Chambers,
  Korzun, Rabinovitch \& Ni]{PNAS2023}
{\sc \au{Gorman, M.~T.}, \au{Rubio, J.~S.}, \au{Diaz-Lopez, M.~X.},
  \au{Chambers, W.~A.}, \au{Korzun, A.~M.}, \au{Rabinovitch, J.} \& \au{Ni,
  R.}} \yr{2023}  \at{Scaling laws of plume-induced granular cratering}.
  \jt{PNAS Nexus}  \bvol{2}~(9),  \pg{pgad300}.

\bibitem[Guleria \& Patil(2020)]{guleria2020}
{\sc \au{Guleria, S.~D.} \& \au{Patil, D.~V.}} \yr{2020}  \at{Experimental
  investigations of crater formation on granular bed subjected to an air-jet
  impingement}.  \jt{Physics of Fluids}  \bvol{32}~(5),  \pg{053309}.

\bibitem[Holsapple \& Schmidt(1982)]{holsapple1982}
{\sc \au{Holsapple, K.~A.} \& \au{Schmidt, R.~M.}} \yr{1982}  \at{On the
  scaling of crater dimensions: 2. impact processes}.  \jt{Journal of
  Geophysical Research: Solid Earth}  \bvol{87}~(B3),  \pg{1849--1870}.

\bibitem[Katsuragi(2010)]{katsuragi2010}
{\sc \au{Katsuragi, H.}} \yr{2010}  \at{Morphology scaling of drop impact onto
  a granular layer}.  \jt{Physical review letters}  \bvol{104}~(21),
  \pg{218001}.

\bibitem[Katsuragi(2016)]{katsuragi2016}
{\sc \au{Katsuragi, H.}} \yr{2016} {\em Physics of soft impact and
  cratering\/}, ,  \vol{vol. 910}.  \publ{1st edn., Lecture Notes in Physics,
  Springer Japan}.

\bibitem[Kuang {\em et~al.\/}(2013)Kuang, LaMarche, Curtis \& Yu]{kuang2013}
{\sc \au{Kuang, S.~B.}, \au{LaMarche, C.~Q.}, \au{Curtis, J.~S.} \& \au{Yu,
  A.~B.}} \yr{2013}  \at{Discrete particle simulation of jet-induced cratering
  of a granular bed}.  \jt{Powder Technol.}  \bvol{239},  \pg{319--336}.

\bibitem[LaMarche \& Curtis(2015)]{lamarche2015}
{\sc \au{LaMarche, C.~Q.} \& \au{Curtis, J.~S.}} \yr{2015}  \at{Cratering of a
  particle bed by a subsonic turbulent jet: Effect of particle shape, size and
  density}.  \jt{Chemical Engineering Science}  \bvol{138},  \pg{432--445}.

\bibitem[Lane {\em et~al.\/}(2010)Lane, Metzger, Clements \&
  Immer]{lane2010cratering}
{\sc \au{Lane, J.~E.}, \au{Metzger, P.~T.}, \au{Clements, S.} \& \au{Immer,
  C.~D.}} \yr{2010}  \at{Cratering and blowing soil by rocket engines during
  lunar landings}.  \bt{In {\em Lunar Settlements\/}},  \pg{pp. 569--594}.
  \publ{CRC Press}.

\bibitem[Launder \& Rodi(1979)]{launder1979}
{\sc \au{Launder, B.~E.} \& \au{Rodi, W.}} \yr{1979}  \at{The turbulent wall
  jet}.  \jt{Progress in Aerospace Sciences}  \bvol{19},  \pg{81--128}.

\bibitem[Launder \& Rodi(1983)]{launder1983}
{\sc \au{Launder, B.~E.} \& \au{Rodi, W.}} \yr{1983}  \at{The turbulent wall
  jet measurements and modeling}.  \jt{Annual review of fluid mechanics}
  \bvol{15}~(1),  \pg{429--459}.

\bibitem[Lohse {\em et~al.\/}(2004)Lohse, Bergmann, Mikkelsen, Zeilstra, Van
  Der~Meer, Versluis, Van Der~Weele, van~der Hoef \& Kuipers]{lohse2004}
{\sc \au{Lohse, D.}, \au{Bergmann, R.}, \au{Mikkelsen, R.}, \au{Zeilstra, C.},
  \au{Van Der~Meer, D.}, \au{Versluis, M.}, \au{Van Der~Weele, K.}, \au{van~der
  Hoef, M.} \& \au{Kuipers, H.}} \yr{2004}  \at{Impact on soft sand: void
  collapse and jet formation}.  \jt{Physical review letters}  \bvol{93}~(19),
  \pg{198003}.

\bibitem[Maruoka(2023)]{maruoka2023}
{\sc \au{Maruoka, H.}} \yr{2023}  \at{A framework for crossover of scaling law
  as a self-similar solution: dynamical impact of viscoelastic board}.  \jt{The
  European Physical Journal E}  \bvol{46}~(5),  \pg{35}.

\bibitem[Metzger(2024{\natexlab{{\em a\/}}})]{metzger2024a}
{\sc \au{Metzger, P.~T.}} \yr{2024{\natexlab{{\em a\/}}}}  \at{Erosion rate of
  lunar soil under a landing rocket, part 1: Identifying the rate-limiting
  physics}.  \jt{Icarus}  \bvol{417},  \pg{116136}.

\bibitem[Metzger(2024{\natexlab{{\em b\/}}})]{metzger2024b}
{\sc \au{Metzger, P.~T.}} \yr{2024{\natexlab{{\em b\/}}}}  \at{Erosion rate of
  lunar soil under a landing rocket, part 2: Benchmarking and predictions}.
  \jt{Icarus}  \bvol{417},  \pg{116135}.

\bibitem[Metzger {\em et~al.\/}(2009)Metzger, Latta~III, Schuler \&
  Immer]{metzger2009}
{\sc \au{Metzger, P.~T.}, \au{Latta~III, R.~C.}, \au{Schuler, J.~M.} \&
  \au{Immer, C.~D.}} \yr{2009} Craters formed in granular beds by impinging
  jets of gas.  \bt{In {\em AIP Conference Proceedings\/}}, ,  \vol{vol. 1145},
   \pg{pp. 767--770}. American Institute of Physics.

\bibitem[Metzger {\em et~al.\/}(2011)Metzger, Smith \& Lane]{metzger2011}
{\sc \au{Metzger, P.~T.}, \au{Smith, J.} \& \au{Lane, J.~E.}} \yr{2011}
  \at{Phenomenology of soil erosion due to rocket exhaust on the moon and the
  mauna kea lunar test site}.  \jt{Journal of Geophysical Research: Planets}
  \bvol{116}~(E6).

\bibitem[Nefzaoui \& Skurtys(2012)]{nefzaoui2012}
{\sc \au{Nefzaoui, E.} \& \au{Skurtys, O.}} \yr{2012}  \at{Impact of a liquid
  drop on a granular medium: Inertia, viscosity and surface tension effects on
  the drop deformation}.  \jt{Experimental Thermal and Fluid Science}
  \bvol{41},  \pg{43--50}.

\bibitem[Prieur {\em et~al.\/}(2017)Prieur, Rolf, Luther, W{\"u}nnemann, Xiao
  \& Werner]{prieur2017}
{\sc \au{Prieur, N.~C.}, \au{Rolf, T.}, \au{Luther, R.}, \au{W{\"u}nnemann,
  K.}, \au{Xiao, Z.} \& \au{Werner, S.~C.}} \yr{2017}  \at{The effect of target
  properties on transient crater scaling for simple craters}.  \jt{Journal of
  Geophysical Research: Planets}  \bvol{122}~(8),  \pg{1704--1726}.

\bibitem[Rajaratnam \& Beltaos(1977)]{rajaratnam1977}
{\sc \au{Rajaratnam, N.} \& \au{Beltaos, S.}} \yr{1977}  \at{Erosion by
  impinging circular turbulent jets}.  \jt{Journal of the Hydraulics Division}
  \bvol{103}~(10),  \pg{1191--1205}.

\bibitem[Schlichting \& Gersten(1979)]{schlichting1979}
{\sc \au{Schlichting, Hermann} \& \au{Gersten, K}} \yr{1979}
  \at{Boundary-layer theory 7th ed}.  \bt{In {\em Kestin J. Chs. 14 and 20\/}}.
   \publ{McGraw-hill New York}.

\bibitem[Scott \& Ko(1968)]{scott1968}
{\sc \au{Scott, R.~F.} \& \au{Ko, H.~Y.}} \yr{1968}  \at{Transient
  rocket-engine gas flow in soil.}  \jt{AIAA Journal}  \bvol{6}~(2),
  \pg{258--264}.

\bibitem[Toigo \& Richardson(2003)]{toigo2003}
{\sc \au{Toigo, A.~D.} \& \au{Richardson, M.~I.}} \yr{2003}  \at{Meteorology of
  proposed mars exploration rover landing sites}.  \jt{Journal of Geophysical
  Research: Planets}  \bvol{108}~(E12).

\bibitem[Uehara {\em et~al.\/}(2003)Uehara, Ambroso, Ojha \&
  Durian]{uehara2003}
{\sc \au{Uehara, J.~S.}, \au{Ambroso, M.~A.}, \au{Ojha, R.~P.} \& \au{Durian,
  D.~J.}} \yr{2003}  \at{Low-speed impact craters in loose granular media}.
  \jt{Physical Review Letters}  \bvol{90}~(19),  \pg{194301}.

\bibitem[Van Der~Meer(2017)]{van2017}
{\sc \au{Van Der~Meer, D.}} \yr{2017}  \at{Impact on granular beds}.
  \jt{Annual review of fluid mechanics}  \bvol{49},  \pg{463--484}.

\bibitem[Witze(2023)]{witze2023}
{\sc \au{Witze, A.}} \yr{2023}  \at{Moon mission failure: why is it so hard to
  pull off a lunar landing?}  \jt{Nature} {URL:
  https://doi.org/10.1038/d41586-023-01454-7}.

\bibitem[Yamamoto {\em et~al.\/}(2017)Yamamoto, Hasegawa, Suzuki \&
  Matsunaga]{yamamoto2017}
{\sc \au{Yamamoto, S.}, \au{Hasegawa, S.}, \au{Suzuki, A.~I.} \& \au{Matsunaga,
  T.}} \yr{2017}  \at{Impact velocity dependence of transient cratering
  growth}.  \jt{Journal of Geophysical Research: Planets}  \bvol{122}~(5),
  \pg{1077--1089}.

\bibitem[Yue(2001)]{yue2001}
{\sc \au{Yue, Z.}} \yr{2001}  \at{Air jets in ventilation applications}.
  \jt{Building Service Engineering, Royal Institute of Technology, Bulletin}
  ~(55).

\bibitem[Zhao {\em et~al.\/}(2015{\natexlab{{\em a\/}}})Zhao, Zhang, Tjugito \&
  Cheng]{zhao2015}
{\sc \au{Zhao, R.}, \au{Zhang, Q.}, \au{Tjugito, H.} \& \au{Cheng, X.}}
  \yr{2015{\natexlab{{\em a\/}}}}  \at{Granular impact cratering by liquid
  drops: understanding raindrop imprints through an analogy to asteroid
  strikes}.  \jt{Proceedings of the National Academy of Sciences}
  \bvol{112}~(2),  \pg{342--347}.

\bibitem[Zhao {\em et~al.\/}(2015{\natexlab{{\em b\/}}})Zhao, de~Jong \&
  van~der Meer]{zhaomeer2015}
{\sc \au{Zhao, S.~C.}, \au{de~Jong, R.} \& \au{van~der Meer, D.}}
  \yr{2015{\natexlab{{\em b\/}}}}  \at{Raindrop impact on sand: a dynamic
  explanation of crater morphologies}.  \jt{Soft Matter}  \bvol{11}~(33),
  \pg{6562--6568}.

\bibitem[Zhao {\em et~al.\/}(2017)Zhao, de~Jong \& van~der Meer]{zhao2017}
{\sc \au{Zhao, S.~C.}, \au{de~Jong, R.} \& \au{van~der Meer, D.}} \yr{2017}
  \at{Liquid-grain mixing suppresses droplet spreading and splashing during
  impact}.  \jt{Physical review letters}  \bvol{118}~(5),  \pg{054502}.

\bibitem[Zhao {\em et~al.\/}(2013)Zhao, Zhao \& Liu]{zhao2013}
{\sc \au{Zhao, Z.}, \au{Zhao, J.} \& \au{Liu, H.}} \yr{2013}  \at{Landing
  dynamic and key parameter estimations of a landing mechanism to asteroid with
  soft surface}.  \jt{International Journal of Aeronautical and Space Sciences}
   \bvol{14}~(3),  \pg{237--246}.

\end{thebibliography}

\end{document}